\documentstyle[preprint,aps,epsfig]{revtex}
\tighten
\begin{document}
\preprint{Preprint Numbers: \parbox[t]{45mm}{MZ-TH/99-06}}
\title{Strong and radiative decays of heavy flavored baryons}
\author{M. A. Ivanov \footnotemark[1],
J. G. K\"{o}rner \footnotemark[2],
V. E. Lyubovitskij  \footnotemark[1]\footnotemark[3] and
A. G. Rusetsky \footnotemark[1]\footnotemark[4]\footnotemark[5]
\vspace*{0.4\baselineskip}}
\address{
\footnotemark[1] Bogoliubov Laboratory of Theoretical Physics,\\
Joint Institute for Nuclear Research, 141980 Dubna, Russia
\vspace*{0.2\baselineskip}\\
\footnotemark[2]
Johannes Gutenberg-Universit\"{a}t, \\
Institut f\"{u}r Physik, D-55099 Mainz, Germany
\vspace*{0.2\baselineskip}\\
\footnotemark[3]
Department of Physics, \\
Tomsk State University,
634050 Tomsk, Russia
\vspace*{0.2\baselineskip}\\
\footnotemark[4]
Institute for Theoretical Physics, \\
University of Bern, Sidlerstrasse 5, CH-3012, Bern, Switzerland
\vspace*{0.2\baselineskip}\\
\footnotemark[5]
HEPI, Tbilisi State University, 380086 Tbilisi, Georgia
\vspace*{0.3\baselineskip}}
\date
{Pacs Numbers:
12.39.Ki, 12.60.Rc, 13.30.Eg, 13.30.Hq, 14.20.Lq, 14.20.Mr
}
\maketitle
\begin{abstract}
We analyze strong one-pion and radiative one-photon decays of heavy
flavored baryons within a relativistic three-quark model.
Employing the same parameters as were used for the description of
the semileptonic decays of heavy baryons, we calculate the couplings
of one-pion and one-photon transitions of both ground and excited
heavy baryon states. We predict the decay rates for all
relevant decay modes and compare them with experimental data when
available and with the results of other model calculations.
\end{abstract}
%

\section{Introduction}

The last decade has seen significant experimental progress in charm baryon
physics. Most of the ground state baryons containing one c-quark have now been
established \cite{PDG}. Their classification in the standard quark model is
based on three different spin configurations: the lowest lying configuration
has $J^P=\frac{1}{2}^+$ ($\Lambda_c$, $\Xi_c$) with the two light
quarks in the antisymmetric spin 0 configuration, the next higher
configuration $J^P=\frac{1}{2}^+$ ($\Sigma_c$, $\Xi'_c$, $\Omega_c$) and
the highest configuration $J^P=\frac{3}{2}^+$ ($\Sigma^*_c$, $\Xi^*_c$,
$\Omega^*_c$) where the two light quarks are in the symmetric spin 1
configuration, coupling down and up to $\frac{1}{2}^+$ and $\frac{3}{2}^+$,
respectively, with the heavy quark.
Some of their  decay modes have been seen and have been used for the
determination of their masses (see \cite{PDG}).

The CLEO Coll. \cite{CLEO1} has presented
evidence for a pair of excited charm baryons, one decaying into
$\Lambda_c^+\pi^+$ with a mass difference $M(\Lambda_c^+\pi^+)-M(\Lambda_c^+)$
of $234.5\pm 1.1\pm 0.8$ MeV and a width of $17.9^{+3.8}_{-3.2}\pm 4.0$ MeV,
and the other into $\Lambda_c^+\pi^-$ with a mass difference
$M(\Lambda_c^+\pi^-)-M(\Lambda_c^+)$ of $232\pm 1.0\pm 0.8$ MeV and a width
of $13.0^{+3.7}_{-3.0}\pm 4.0$ MeV. The CLEO Coll. interpreted these data
as evidence for the $\Sigma^{*++}_c$ and $\Sigma^{*0}_c$,
the spin $\frac{3}{2}^+$ excitations of the $\Sigma_c$ baryons.

Earlier, the CLEO Coll. \cite{CLEO2} has also reported
evidence for a pair of excited charmed baryons, one decaying into
$\Xi_c^+\pi^-$ with a mass difference $M(\Xi_c^+\pi^-)-M(\Xi_c^+)$
of $178.2\pm 0.5\pm 1.0$ MeV and a width of $<5.5$ MeV,
and the other \cite{CLEO3} into $\Xi_c^0\pi^+$ with a mass difference
$M(\Xi_c^0\pi^+)-M(\Xi_c^0)$ of $174.3 \pm 0.5\pm 1.0$ MeV and a width
of $<3.1$ MeV. They interpreted these data as evidence
for the $\Xi^{*0}_c$ and $\Xi^{*+}_c$, the spin $\frac{3}{2}^+$
excitations of the $\Xi_c$ baryons.

Recently, the CLEO Coll. \cite{CLEO4} has reported on the
observation of two narrow states in the decay modes $\Xi_c^+\gamma$ and
$\Xi_c^0\gamma$. They have been interpreted as the $\Xi_c'$ states.
The mass differences $M(\Xi_c^{+\prime})-M(\Xi_c^{+})$ and
$M(\Xi_c^{0\prime})-M(\Xi_c^{0})$ have been measured to be
$107.8\pm 1.7\pm 2.5$ MeV and $107.0\pm 1.4\pm 2.5$ MeV, respectively.
Except for the $\Omega_c^*$ and some of the isospin partners of the
charm baryon ground states most of the ground state charm baryons are now
well established.

Three collaborations (ARGUS \cite{ARGUS}, E687 \cite{E687} and
CLEO \cite{CLEO5}) have seen a doublet of particles decaying into
$\Lambda_c^+\pi^+\pi^-$. They have been interpreted as the lowest lying
orbitally excited states of the $\Lambda_c^+$: $\Lambda_{c1}^+(2593)$
with $J^P=\frac{1}{2}^-$ and $\Lambda_{c1}^{*+}(2625)$ with
$J^P=\frac{3}{2}^-$. The PDG average for the decay width of the
$\Lambda_{c1}^+(2593)$ is $3.6^{+2.0}_{-1.3}$ MeV. The branching fractions
into the $\Lambda_c^+\pi^+\pi^-$ and the $\Sigma_c^{++}(2455)\pi^-$
and $\Sigma_c^{0}(2455)\pi^+$ modes are estimated to be $\approx 67\%$,
$24\pm 7\%$ and $18\pm 7\%$, respectively. For the decay width of the
$\Lambda_c^+(2625)$ there exists only an upper limit ($<1.9$ MeV) and branching
ratios \cite{PDG}, \cite{ARGUS}-\cite{CLEO5}.

The CLEO Coll. \cite{CLEO6} has reported preliminary evidence for a new
charm baryon decaying into  $\Xi_c^+\pi^+\pi^-$ via the intermediate
$\Xi_c^{*0}$-state. The measured mass difference is given by
$M(\Xi_c^+\pi^+\pi^-)-M(\Xi_c^+)=349.4\pm 0.7\pm 1.0$ MeV, and its width is
$\Gamma<2.4$ MeV. This new particle was interpreted as the $\Xi^{*+}_{c1}$
with $J^P=\frac{3}{2}^-$, the charm-strange partner
of the $\Lambda_{c1}^{*+}(2625)$.

The classification and the decay properties of ground and excited state
charm and bottom baryons have been reviewed in \cite{KKP}. The analysis
was based on the heavy quark limit given by the leading order in the
$1/m_Q$-expansion. In the heavy quark limit the dynamics of the heavy and
light quarks decouple leading to a number of model independent relations
between various decay modes of the heavy baryons.
Further relations between physical observables can be obtained if one
assumes additional symmetries for the light quark system.
For instance, by using a constituent quark model picture for the light
diquark system with an underlying $SU(2N_f)\otimes O(3)$ symmetry,
a number of further relations were derived in \cite{HKLT} for the form
factors in semileptonic $b-c$ transitions.

The subject of this paper mostly concerns the one-pion and one-photon decays
of ground state charm baryons as well as those of the lowest-lying p-wave
states. We shall also determine the one-photon decays of
the orbital excitations of bottom baryons. The analysis of these transitions
provides an excellent laboratory for tests of heavy quark symmetry predictions
on the one hand and tests of the soft dynamics of the light-side one-pion
(photon) diquark transitions on the other hand. In the heavy
quark limit the pion and photon are emitted from the light diquark system
while the heavy quark is unaffected by the emission process.

Single pion transitions of charm baryons were analyzed before in \cite{KKP,HKT}
by again using the constituent quark model picture for the light
diquark system with its underlying $SU(2N_f)\otimes O(3)$ symmetry.
Using this symmetry one significantly reduces the number of
independent coupling factors \cite{KKP,HKT}.

A similar constituent quark model approach has been employed in \cite{PY}
to establish relationships
among the coupling constants characterizing the decays of s-wave and p-wave
heavy baryons.  The results of the work \cite{PY} are
the same as in \cite{HKT} since both approaches are based on the same
constituent quark model picture.

Light-front (LF) quark model functions with a factorized
harmonic-oscillator transverse momentum component and a longitudinal
component given by a $\delta$-function have been constructed in \cite{TOK}.
They have been employed to calculate the strong couplings for
$\Sigma_c\to\Lambda_c\pi$,
$\Lambda_{c1}\to\Sigma_c\pi$, and $\Lambda_{c1}^*\to\Sigma_c\pi$  decay modes
which correspond to P-wave, S-wave and D-wave transitions, respectively.

The flavor-spin symmetry of the heavy quarks and the spontaneously broken
$SU(3)_L\otimes SU(3)_R$ chiral symmetry of the light quarks were exploited
to describe the interactions of heavy mesons and heavy baryons with
the $\pi$, $K$, and $\eta$ mesons considered as the Goldstone bosons
\cite{Yan}. This approach contains three free parameters and was applied
to strong and semileptonic decays of heavy hadrons. The radiative decays of
heavy mesons and heavy baryons were also studied within this formalism in
\cite{Cheng1}. The two orbital excitations of the $\Lambda_c$ have been
analyzed within this approach which is referred to as the
heavy hadron chiral perturbation theory (HHCPT) by Cho \cite{Cho}
(see also \cite{PY}).
Information on one of the three free parameters of HHCPT in these decays was
obtained from the radiative decay $\Xi_c^{\prime * 0}\to \Xi_c^0\gamma$
by Cheng \cite{Cheng2}. Pion transitions of the ground states and some excited
p-wave states of heavy baryons
were studied within HHCPT in \cite{HDH}. The coupling constants were
estimated using the experimental data on the pion decays of strange
$\Lambda$ and $\Sigma$ baryons treating the $s$-quark as static.

The radiative decays of some excited heavy baryon states have been calculated
in the so-called bound state picture \cite{Chow} motivated by the large
$N_c$ limit. It was shown that the $\Lambda_{c1}(2593)\to\Lambda_c\gamma$
and $\Lambda_{c1}^*(2625)\to\Lambda_c\gamma$ decays are severely suppressed
whereas the $\Lambda_{b1}^*(5900)\to\Lambda_b\gamma$ mode possibly
dominates over the strong decay mode.

Some aspects of the phenomenology of new baryons with charm and strangeness
were discussed in \cite{CF}. The authors of Ref. \cite{CF} estimated the
expected width of the excited state $\Xi_{c1}(\frac{3}{2})^+$
of the $\Xi_c$ family using the experimental value for the width of
the $\Lambda_{c1}(\frac{1}{2})^+$ within the $SU(3)$ flavor symmetry limit.
Also it was pointed out that one should search for the $\Omega_{c2}^*$ baryon
in the $\Xi_c K$ decay channel.

The ratio of electric quadrupole (E2) and magnetic dipole (M1) components
for the decay $\Sigma_c^*\to\Lambda_c\gamma$ has been computed by
Savage \cite{Savage} within the HHCPT approach. The result was that the
$1/m_c$ suppression of the E2 amplitude is compensated for by a small energy
denominator arising from the infrared behavior of pion loop graphs in chiral
perturbation theory. This leads to a $E2/M1$ ratio of the order of a few
percent depending among others on
the $\Sigma_c^* - \Sigma_c$ spin symmetry breaking mass difference.
The corresponding ratio in the bottom baryon sector is smaller by a factor
of $\sim M_c/M_b$.

The couplings in the HHCPT Lagrangian have been estimated from QCD sum rules
in an external axial field \cite {GY}.
The radiative decays of heavy baryons were studied with the light cone QCD
sum rules  in the leading order of HQET in \cite{ZD}.

All approaches considered above exploit the symmetry of the heavy quark limit
with some additional assumptions on the structure of the light quark
system without employing any dynamical scheme for the composite structure of
hadrons.

A relativistic quark model of hadrons \cite{EI}-\cite{ILL} has been developed
some time ago to describe physical observables in the low-energy domain.
The model is based on an effective hadron-quark Lagrangian and the
compositeness condition $Z_H=0$  where $Z_H$ is a wave function renormalization
constant of the hadron \cite{SW}.
This approach has been employed to give a unified description of
leptonic and semileptonic decays of heavy mesons in \cite{IS}.

Recently, this model called the Relativistic Three-Quark Model (RTQM) has
been applied to calculate physical observables in the decays of heavy
baryons \cite{ILKK}-\cite{IKL}. All results of the RTQM model are expressed
through a few model parameters: the masses of the light quarks, the mass
differences of the heavy baryon and heavy quark, and the size of the Gaussian
distribution of constituents inside the hadron.
The exclusive semileptonic and nonleptonic decays of
charmed and bottom baryons have been considered within the RTQM
\cite{ILKK,IKLR1}. Preliminary results for the strong and radiative decays
of heavy baryons have been reported in \cite{IKLR2,IKL}.
In this paper  we extend this application by reporting the simultaneous
calculation of a range of one-pion and one-photon transitions
between heavy baryons.

Our article is divided into five sections with a single appendix.
In Sec. II we give some necessary background material on the RTQM and discuss
features of gauging the interaction Lagrangian in this approach.
The calculation of the matrix elements of the one-pion and one-photon
transitions between heavy baryons is given Sec. III.
In Sec. IV we present our numerical results on the
heavy baryon observables in their strong and radiative transitions.
We compare our results with available experimental data
and with the results of some other model approaches.
We make some concluding remarks in Sec. V. The Appendix contains a detailed
discussion of the gauge invariance of radiative transitions in the RTQM.

\section{Relativistic Three-Quark Model}

A detailed description of the Relativistic Three-Quark Model can be found in
Refs. \cite{ILL,ILKK}. Here we will only give the necessary background
material
needed for the description of two-body strong and electromagnetic decays
of heavy flavored baryons. We will discuss the interaction Lagrangian
describing the coupling of heavy baryons with their constituent quarks and
its gauging. We also specify the form of the light and heavy quark propagators.

The Lagrangian describing the coupling of a heavy baryon to its
constituent light and heavy quarks considerably simplifies in the heavy quark
limit (see \cite{ILKK}). One has
\begin{eqnarray}\label{LHB_int}
{\cal L}_{B_Q}^{\rm int}(x)&=&g_{B_Q}\bar B_Q(x)
\Gamma_1 Q^a(x)\int\hspace*{-0.1cm} d\xi_1\hspace*{-0.1cm}
\int\hspace*{-0.1cm} d\xi_2
F_B(\xi_1^2+\xi_2^2)\\ \label{int_bar}
&\times&q^b(x+3\xi_1-\sqrt{3}\xi_2)C\Gamma_2\lambda_{B_Q}
q^c(x+3\xi_1+\sqrt{3}\xi_2)\varepsilon^{abc}+{\rm h.c.}\nonumber\\
F_B(\xi_1^2+\xi_2^2)&=&
\int\hspace*{-0.1cm}\frac{d^4k_1}{(2\pi)^4} \hspace*{-0.1cm}
\int\hspace*{-0.1cm}\frac{d^4k_2}{(2\pi)^4} \hspace*{0.1cm}
e^{ik_1\xi_1+ik_2\xi_2}
\tilde F_B\biggl\{\frac{[k_1^2+k_2^2]}{\Lambda_B^2}\biggr\}
\nonumber
\end{eqnarray}
The interaction Lagrangian of the pion with its constituent light quarks is
given by
\begin{eqnarray}\label{int_pi}
{\cal L}_\pi^{\rm int}(x)=\frac{ig_\pi}{\sqrt{2}}\vec\pi(x)\int
\hspace*{-0.1cm} d\xi F_\pi(\xi^2)\bar q(x+\xi/2)
\gamma^5\vec\lambda_\pi q(x-\xi/2)
\end{eqnarray}
where
\begin{eqnarray}
F_\pi(\xi^2)=\int\hspace*{-0.1cm}\frac{d^4k}{(2\pi)^4}
e^{ik\xi} \tilde F_\pi\biggl\{\frac{k^2}{\Lambda_\pi^2}\biggr\}
\nonumber
\end{eqnarray}
Here $\Gamma_i$ and $\lambda_{B_Q}$ are spinor and flavor matrices
which define the quantum numbers of the relevant three-quark currents.
They are listed in Table I. The square brackets $[...]$ and curly brackets
$\{...\}$ denote antisymmetric and symmetric flavor and spin combinations of
the light degrees of freedom\footnote{We use the following 
notation for excited heavy $P$-wave baryons: \\
i) $\Lambda_{c1; S}$, $\Lambda_{c1; S}^*$,
$\Sigma_{c0; S}$ and
$\Xi_{c1; S}^*$ are symmetric under the exchange of the momenta 
of the two light quarks 
(called $K$-states in \cite{KKP,HKLT,HKT}); 
ii) $\Sigma_{c1; A}$ and $\Sigma_{c1; A}^*$ are antisymmetric under the 
exchange of the momenta of the two light quarks
(called $k$-states in \cite{KKP,HKLT,HKT}).}. 
The coupling strength of the respective
hadrons with their constituent quarks are denoted by the coupling constants
$g_{B_Q}$ and $g_\pi$. The parameters
$\Lambda_B$ and $\Lambda_\pi$ define the size of the distributions of light
quarks inside the heavy baryon and pion, respectively.
The baryon  parameter $\Lambda_B$ is chosen to be
the same for charm and bottom baryons to provide the correct
normalization of the baryonic Isgur-Wise function  in the heavy
quark limit \cite{ILKK}.

The gauging of the nonlocal interaction Lagrangian Eq.(\ref{int_bar})
can be done by using a path-independent formalism based on the
path-independent definition \cite{Mandelstam} of the derivative of the path
integral. The details may be found in \cite{Terning} and also
in \cite{ILL} where this formalism was employed to calculate
the nucleon electromagnetic form factors in the quark model in a gauge
invariant way. In order to make the Lagrangian
Eq.(\ref{LHB_int}) gauge invariant in the presence of an electromagnetic
field $A_\mu(x)$ the time-ordered P-exponent
\begin{equation}\label{exp}
P\exp\biggl\{ieQ\int\limits_x^y dz_\mu A^\mu(z)\biggr\}
\end{equation}
($Q=\rm diag\{2/3,-1/3,-1/3\}$ is the charge matrix)
is attached to each light quark field $q(y)$. We then have
\begin{eqnarray}\label{Lagr_GI}
{\cal L}_{B_Q}^{\rm int, e.m.}(x)&=&g_{B_Q}\bar B_Q(x)\Gamma_1 Q^a(x)
\int d^4\xi_1\int d^4\xi_2F_B(\xi_1^2+\xi_2^2)\\
&\times&\exp(ieQ\int\limits_x^{x+3\xi_1-\sqrt{3}\xi_2} dz_\mu A^\mu(z))
q^b(x+3\xi_1-\sqrt{3}\xi_2 )C\Gamma_2\lambda_{B_Q}\nonumber\\
&\times&\exp(ieQ\int\limits_x^{x+3\xi_1+\sqrt{3}\xi_2} dz_\mu A^\mu(z))
q^c(x+3\xi_1+\sqrt{3}\xi_2)\varepsilon^{abc}+{\rm h.c.}\nonumber
\end{eqnarray}
The Lagrangian Eq. (\ref{Lagr_GI}) generates nonlocal vertices which involve
the heavy baryons, photons and the light and heavy  quarks. The gauge
formalism of Mandelstam is based on the definition of the derivative
of the path integral
\begin{equation}\label{path}
I(y,x,P)=\int\limits_x^y dz_\mu A^\mu(z)
\end{equation}
where P is the path taken from $x$ to $y$. When calculating Feynman diagrams
the derivative of  $I(y,x,P)$ is defined such that

$$
\frac{\partial}{\partial y^\mu}I(y,x,P)=A_\mu(y).
$$
This means that the field $A_\mu(y)$ does not depend on the path used
in the definition of the line integral. The Mandelstam prescription makes
all matrix elements involving a photon gauge-invariant and path-independent.

The free  quark Lagrangian and the interaction Lagrangian for transitions
involving the orbital excited states (P-states, etc.) are gauged by
the standard minimal substitution, i.e. for the derivative coupling appearing
in the $P$-states one has  to replace $\partial_\mu\to\partial_\mu+ieA_\mu$.

Now we discuss the implementation of gauge invariance in the one-photon
transitions of heavy baryons.
There are several relevant diagrams:
i) the triangle diagrams in Fig. 2a and 2b  with a photon emitted
by the heavy and light quark, respectively
ii) the contact interaction-type diagrams in Fig. 3 with a photon emitted
by one of the two nonlocal baryon-quark vertices
and
iii) the pole diagrams in Fig. 4 with a photon emitted by one of the two heavy
baryons.

The contributions coming from the triangle diagram in Fig. 2a
and the pole diagrams in Fig. 4 are nonleading in $1/m_Q$ and vanish
in the heavy quark limit.
The contact interaction-type diagrams in Fig. 3 are important to reproduce
the  Ward-Takahashi identities for the connected Green functions
(see, for example,  Ref. \cite{ILL}). Their contributions
to the matrix element of the one-photon transitions of ground state charm
baryons are nonleading in the heavy mass expansion,
at least when the photon is on its mass shell $(q^2=0)$.
However, they  contribute to the matrix elements
of one-photon transitions of the orbitally excited charm baryon states
$B_{c1}\to B_c\gamma$ and $B_{c1}^*\to B_c\gamma$ where
$B_c=\Lambda_c$ or $\Xi_c$. Formally the contact interaction type
contributions in this case are of the order
$O(qv)\approx O(0)$ ($qv=(m_i^2-m_f^2)/(2m_i) \approx 0$) and one would
naively assume that they vanish in the heavy quark limit.
However, contrary to the ground state transitions, the excited state
transitions $B_{c1}\to B_c\gamma$ and $B_{c1}^*\to B_c\gamma$ are of the
same order $Q(qv)$ and thus the contact interaction-type contribution must
be kept in this case. In the Appendix we provide a detailed discussion of
the gauge-invariance structure of the matrix
elements for one-photon transitions of excited charm
baryons\footnote{In our previous analysis of we neglected the contact
interaction-type diagrams of the $P$-wave heavy baryon
transitions \cite{IKL}. Their
contribution to the invariant matrix element is very small of order 3-4\%.
We demonstrate in the Appendix that the contact interaction-type
contribution is needed to reproduce the correct gauge invariance structure
of the photon transitions.}.

For radiative transitions involving P-wave states there are additional contact
interaction-type diagrams which contribute to the leading order in the
heavy quark expansion. They result from the minimal substitution prescription
for the derivatives acting on the excited
heavy baryon fields in the heavy baryon-three-quark interaction
Lagrangian (\ref{LHB_int}). In the following we will refer to such
diagrams as "local contact" diagrams in order to distinguish them from the
contact diagrams generated by the gauging of the nonlocal heavy
baryon-three-quark vertex ("nonlocal contact" diagrams).

Thus, in the heavy quark limit the radiative decay of a ground state heavy
baryon is described by the solely triangle diagram in  Fig. 2b with the
photon emitted by the light quark whereas there are additional contributions
coming from both the ``nonlocal contact'' in Fig. 3 and ``local contact''
in Fig. 4 diagrams for the radiative decays of heavy excited baryon states.
For the one-pion transitions one only needs the only triangle diagram
depicted in Fig. 1.

Let us now specify our model parameters. For the light quark propagator with
a constituent mass $m_q$ we shall use the standard form of the free fermion
propagator
\begin{equation}\label{light}
S_q(k)=\frac{1}{m_q-\not\! k}.
\end{equation}
For the heavy quark propagator we shall use the leading term
$S_v(k,\bar\Lambda_{\{q_1q_2\}})$ in the inverse mass expansion of
the free fermion propagator:
\begin{eqnarray}
S_Q(p+k)&=&\frac{1}{m_Q-(\not\! p+\not\! k) }=
S_v(k,\bar\Lambda_{\{q_1q_2\}})
+O\biggl(\frac{1}{m_Q}\biggr),
\nonumber\\
&&\nonumber\\
S_v(k,\bar\Lambda_{\{q_1q_2\}})&=&-\frac{1+\not\! v}
{2(vk+\bar\Lambda_{\{q_1q_2}\})}. \label{heavy}
\end{eqnarray}
where we introduce the mass difference parameter
$\bar\Lambda_{\{q_1q_2\}}=M_{\{Qq_1q_2\}}-m_Q$ which is the difference
between the heavy baryon mass $M_{\{Qq_1q_2\}}\equiv M_{B_Q}$ and the heavy
quark mass $m_Q$. The four-velocity of the heavy quark is denoted by $v$ as
usual.  We shall neglect a possible mass difference between the constituent
$u-$ and $d-$quark and thus take $\bar\Lambda : =  \bar\Lambda_{uu}=
\bar\Lambda_{ud}= \bar\Lambda_{dd}$, $ \bar\Lambda_s : = \bar\Lambda_{us}=
 \bar\Lambda_{ds}$ meaning that there are altogether three independent
mass difference parameters: $ \bar\Lambda$, $ \bar\Lambda_s$, and
$\bar\Lambda_{ss}$.

The vertex functions $F_B$ and $F_\pi$ are arbitrary functions except
that they should fall off sufficiently fast in the ultraviolet region to
render the Feynman diagrams ultraviolet finite.
In principle, their functional forms are calculable from the solutions
of the Bethe-Salpeter equations for the baryon and pion bound states.
For example, in \cite{IKMR} semileptonic meson transitions have been
analyzed using the heavy-quark limit of the Dyson-Schwinger
equation. The results were found to be quite insensitive to the details of
the functional form of the heavy-meson Bethe-Salpeter amplitude.
We will use this observation as a guiding principle and choose simple
Gaussian forms for the vertices $F_B$ and $F_\pi$.
Their Fourier transforms read

\begin{equation}\label{BS}
\tilde F_B(k_1^2+k_2^2)=\exp\left(\frac{k_1^2+k_2^2}{\Lambda_B^2}\right),
\hspace{1cm}
\tilde F_\pi(k^2)=\exp\left(\frac{k^2}{\Lambda_\pi^2}\right).
\end{equation}
where $\Lambda_B$ and $\Lambda_\pi$ characterize the size of the
distributions of the light quarks inside a baryon and pion, respectively.

The coupling constants $g_{B_Q}$ and $g_\pi$ in Eqs. (1) and (2) are calculated
from {\it the compositeness condition} (see, ref.~\cite{EI,SW}), which means
that the renormalization constant of the hadron wave function is set equal to
zero:
$Z_{B_Q}=1-g_{B_Q}^2\Sigma^\prime_{B_Q}(M_{B_Q})=0$ where $\Sigma_{B_Q}$ is
a baryon  mass operator, and similarly for the pion
$Z_\pi=1-g_\pi^2\Pi^\prime_\pi(m^2_\pi)=0$ where $\Pi_\pi$ is
a pion mass operator.

A drawback of our approach is the lack of confinement. To avoid
the appearance of unphysical imaginary parts in the Feynman
diagrams, we shall require that $M_{B_Q}<m_Q+m_{q_1}+m_{q_2}$  which
implies that the mass difference parameter $\bar\Lambda_{q_1q_2}$ is bounded
from above by the requirement
$\bar\Lambda_{q_1q_2}<m_{q_1}+m_{q_2}$.

As mentioned before the masses of the $u$ and the $d$ quarks are set equal
($m_u=m_d=m_q$). The value of $m_q$ is determined from an analysis of
nucleon data: $m_q$=420 MeV \cite{ILL}.
The pion Gaussian size parameter $\Lambda_\pi=$ 1 GeV
is fixed from the description of low-energy pion observables (the coupling
constants $f_\pi$ and $g_{\pi\gamma\gamma}$, and the electromagnetic radii
and form factors of the transitions $\pi\to\pi\gamma$ and
$\pi\to\gamma\gamma^*$)\cite{AIKL}-\cite{IL}.
The parameters $\Lambda_B$, $m_s$, $\bar\Lambda$ are taken from
an analysis of the $\Lambda^+_c\to\Lambda^0+e^+ +\nu_e$ decay data.
A good description of the present average value of the branching ratio
$B(\Lambda_c^+\to\Lambda e^+ \nu_e$) = 2.2 $\%$ can be achieved using
the following values of the parameters $\Lambda_B$, $m_s$ and $\bar\Lambda$:
$\Lambda_B$=1.8 GeV, $m_s$=570 MeV and $\bar\Lambda$=600 MeV.
The values of the  parameters $\bar\Lambda_s$ and $\bar\Lambda_{ss}$
are determined from the heuristic relations
$\bar\Lambda_s = \bar\Lambda + (m_s - m)$
and $\bar\Lambda_{ss} = \bar\Lambda + 2(m_s - m)$, which give
$\bar\Lambda_s$ = 750 MeV and $\bar\Lambda_{ss}$ = 900 MeV. We mention that,
using the same values of $\Lambda_{B_Q}$=1.8 GeV and $\bar\Lambda$=600 MeV,
one obtains a width of $5.4\times 10^{10}s^{-1}$ and a value of
$\rho^2 = 1.4$ for the slope of the Isgur-Wise function in the decay
$\Lambda_b^0\to\Lambda_c^+ e^- \bar\nu_e$ \cite{IKLR2}.
Finally, the mass values of the charm baryon states including current
experimental errors are listed in TABLE I \cite{PDG,CLEO4}.
For the pion masses we take
$m_{\pi^\pm} = $ 139.6 MeV and $m_{\pi^0}$ = 135 MeV~\cite{PDG}.

All dimensional parameters entering the Feynman diagrams are expressed
in units of $\Lambda_B$. The integrals are calculated in the Euclidean
region both for internal and external momenta. The final results are obtained
by analytic continuation of the external momenta to the physical region
after the internal momenta have been integrated out.

\section{Matrix elements of one-pion and one-photon transitions}

We begin our discussion with the one-pion and
one-photon transitions of the ground state baryons. As discussed in Sec. II
in the heavy quark limit they are described by the triangle two-loop diagrams
Fig. 1 and Fig. 2b.

The contribution of the triangle diagram (Fig. 1 and Fig. 2b) to the matrix
element of the one-pion (one-photon) transition
$B^i_Q(p)\to B^f_Q(p^\prime)+X(q)$ has the following form in
the heavy quark limit
\begin{eqnarray}\label{vertex2}
M_{inv, \Delta}^X(B_Q^i\to B_Q^fX)=g_Xg^i_{\rm eff} g^f_{\rm eff}
\cdot \bar u(v^\prime)\Gamma_1^f \frac{(1+\not\! v)}{2}\Gamma_1^iu(v)
\cdot I_{q_1q_2, \Delta}^{if}(v,q)
\end{eqnarray}
where the symbol $X$ denotes a pion or a photon in the final state,
i.e. $X=\pi$ or $\gamma$
and $g_X=g_\pi/\sqrt{2}$ (pion-quark-antiquark coupling) or
$e$ (electron charge);
$g_{\rm eff}=g_{B_Q} \Lambda_B^2/ \sqrt{6}/(8\pi^2)$.
$I_{q_1q_2, \Delta}^{if}(v,q)$ is a two-loop quark integral the form
of which depends on whether one is computing pion or photon transitions.
For the pion one has

\begin{eqnarray}\label{int_pion}
\hspace*{-1cm}I_{q_1q_2, \Delta}^{if}(v,q)|_{X=\pi}&=&
\int\hspace*{-0.1cm}\frac{d^4k_1}{\pi^2i} \hspace*{-0.1cm}
\int\hspace*{-0.1cm}\frac{d^4k_2}{4\pi^2i} \hspace*{0.1cm}
\frac{\tilde F_B(k_1,k_2,q)\tilde F_B(k_1,k_2,0)}
{[-k_1v-\bar\Lambda_{q_1q_2}]}
\Pi_{q_1q_2, \Delta}^\pi(k_1,k_2,q)\\[2mm]
\tilde F_B(k_1,k_2,q)&\equiv&
\tilde F_B\biggl\{-6\biggl[(k_1+q)^2+(k_2-q)^2+(k_1+k_2)^2\biggr]\biggr\}
\nonumber
\end{eqnarray}
Here
$\Pi_{q_1q_2, \Delta}^\pi(k_1,k_2,q)$ is the structure integral
corresponding to the light quark loop:
\begin{eqnarray}\label{Pi1}
\Pi_{q_1q_2, \Delta}^\pi(k_1,k_2,q)=C_{\rm flavor}
\tilde F_\pi\biggl\{-\biggl(k_2-\frac{q}{2}\biggr)^2\biggr\}
{\rm tr}\biggl[\Gamma_2^i S_{q_2}(k_1+k_2)\Gamma_2^f
S_{q_1}(k_2-q)\gamma^5 S_{q_1}(k_2)\biggr]
\end{eqnarray}
where $C_{\rm flavor}={\rm tr}[\lambda_\pi \lambda_{B^i} \lambda_{B^f}]$ is
a trace of flavor matrices and $\Gamma_{1(2)}^i$ and $\Gamma_{1(2)}^f$
are the Dirac matrices of the initial and the final baryons, respectively.

For the photon transition one has

\begin{eqnarray}\label{int_el}
\hspace*{-1cm}I_{q_1q_2, \Delta}^{if}(v,q)|_{X=\gamma}&=&
\varepsilon^*_\mu(q)
\int\hspace*{-0.1cm}\frac{d^4k_1}{\pi^2i} \hspace*{-0.1cm}
\int\hspace*{-0.1cm}\frac{d^4k_2}{4\pi^2i} \hspace*{0.1cm}
\frac{\tilde F_B(k_1,k_2,q)\tilde F_B(k_1,k_2,0)}
{[-k_1v-\bar\Lambda_{q_1q_2}]}
\Pi_{q_1q_2, \Delta}^{\gamma;\mu}(k_1,k_2,q)\\[2mm]
\tilde F_B(k_1,k_2,q)&\equiv&
\tilde F_B\biggl\{-6\biggl[(k_1+q)^2+(k_2-q)^2+(k_1+k_2)^2\biggr]\biggr\}
\nonumber
\end{eqnarray}
\begin{eqnarray}\label{Pi2}
\Pi_{q_1q_2, \Delta}^{\gamma;\mu}(k_1,k_2,q)&=&
Q_{q_2q_2}{\rm tr}\biggl[\Gamma_2^i S_{q_1}(k_1+k_2)
\Gamma_2^fS_{q_2}(k_2-q)\gamma^\mu S_{q_2}(k_2)\biggr]\\
&-&Q_{q_1q_1}{\rm tr}\biggl[\Gamma_2^f S_{q_2}(-k_1-k_2)\Gamma_2^i
S_{q_1}(-k_2)\gamma^\mu S_{q_1}(-k_2+q)\biggr]
\nonumber
\end{eqnarray}
where $\varepsilon^*_\mu(q)$ is the polarization vector of the photon.

Radiative transitions of excited heavy baryon states involve additional
contributions from the "local contact" and "nonlocal contact"
diagrams. In our recent analysis \cite{IKL} we have neglected
the "nonlocal contact" diagrams because their contribution to the
corresponding invariant matrix elements is quite small (of order 3-4\%).
Here we present a complete analysis of radiative transitions of heavy baryons
by taking into account all possible diagrams contributing to these processes in
the heavy quark limit as is necessary to obtain a gauge invariant result.
In the heavy quark limit  the contributions of the
"local contact" diagrams (lcd) and "nonlocal contact" diagrams (ncd) to
the matrix element of the excited heavy baryon transition
$B^i_{Q1}(p)\to B^f_Q(p^\prime)+\gamma$ simplify to
\begin{eqnarray}\label{vertex_add1}
M_{\rm inv, \rm lcd}^\gamma(B_{Q1}^i\to B_Q^f\gamma)=
eg^i_{\rm eff} g^f_{\rm eff}
\cdot \bar u(v^\prime)\Gamma_1^f \frac{(1+\not\! v^\prime)}{2}
\Gamma_1^iu_\nu(v)\cdot I_{q_1q_2, {\rm lcd}}^{if, \nu}(v,q)
\end{eqnarray}
\begin{eqnarray}\label{int_pion10}
\hspace*{-1cm}I_{q_1q_2, {\rm lcd}}^{if, \nu}(v,q)&=& - \varepsilon^*_\nu(q)
\int\hspace*{-0.1cm}\frac{d^4k_1}{\pi^2i} \hspace*{-0.1cm}
\int\hspace*{-0.1cm}\frac{d^4k_2}{4\pi^2i} \hspace*{0.1cm}
\frac{\tilde F_B(k_1,k_2,0)\tilde F_B(k_1,k_2,-q)}
{[-k_1v^\prime-\bar\Lambda_{q_1q_2}]}
\Pi_{q_1q_2}^\gamma(k_1,k_2)
\nonumber
\end{eqnarray}
where
\begin{eqnarray}
\Pi_{q_1q_2}^{\gamma}(k_1,k_2)=
Q_{q_2q_2}{\rm tr}\biggl[\Gamma_2^i S_{q_2}(k_1+k_2)
\Gamma_2^f S_{q_2}(k_2)\biggr]
+Q_{q_1q_1}{\rm tr}\biggl[\Gamma_2^f S_{q_2}(k_1+k_2)\Gamma_2^i
S_{q_1}(k_2)\biggr]
\nonumber
\end{eqnarray}
and
\begin{eqnarray}\label{vertex_add2}
M_{\rm inv, ncd}^\gamma(B_{Q1}^i\to B_Q^f\gamma)&=&
eg^i_{\rm eff} g^f_{\rm eff}
\cdot \bar u(v^\prime)\Gamma_1^f \biggl[\frac{(1+\not\! v)}{2}
I_{q_1q_2, {\rm ncd}}^{if, \nu; L}(v,q) \nonumber\\
&+&\frac{(1+\not\! v^\prime)}{2}
I_{q_1q_2, {\rm ncd}}^{if, \nu; R}(v,q)\biggr]\Gamma_1^i u_\nu(v)
\end{eqnarray}
\begin{eqnarray}\label{int_pion20}
\hspace*{-1cm}I_{q_1q_2, {\rm ncd}}^{if, \nu; L}(v,q)&=&\varepsilon^*_\mu(q)
\int\hspace*{-0.1cm}\frac{d^4k_1}{\pi^2i} \hspace*{-0.1cm}
\int\hspace*{-0.1cm}\frac{d^4k_2}{4\pi^2i} \hspace*{0.1cm}
(k_1-k_2+q)^\mu k_1^\nu \frac{\tilde F_B(k_1,k_2,0)}
{[-k_1v-\bar\Lambda_{q_1q_2}]}\nonumber\\
&\times&\frac{\tilde F_B(k_1,k_2,0)-\tilde F_B(k_1,k_2,q)}
{(k_1-k_2)q+q^2}\Pi_{q_1q_2}^\gamma(k_1,k_2)\\[2mm]
\nonumber\\
\hspace*{-1cm}I_{q_1q_2, {\rm ncd}}^{if, \nu; R}(v,q)&=&\varepsilon^*_\mu(q)
\int\hspace*{-0.1cm}\frac{d^4k_1}{\pi^2i} \hspace*{-0.1cm}
\int\hspace*{-0.1cm}\frac{d^4k_2}{4\pi^2i} \hspace*{0.1cm}
(k_1-k_2-q)^\mu k_1^\nu \frac{\tilde F_B(k_1,k_2,0)}
{[-k_1v^\prime-\bar\Lambda_{q_1q_2}]}\nonumber\\
&\times&\frac{\tilde F_B(k_1,k_2,0)-\tilde F_B(k_1,k_2,-q)}
{(k_2-k_1)q+q^2}\Pi_{q_1q_2}^\gamma(k_1,k_2)\\[2mm]
\nonumber
\end{eqnarray}

As an illustration of our calculational procedure we evaluate a typical
integral corresponding to a one-photon transition (the same integral
for the one-pion transition can be found in Ref. \cite{IKLR2})
\begin{eqnarray}
R^{if; \mu}_{q_1q_2}(v,q) =
\int\hspace*{-0.1cm}\frac{d^4k_1}{\pi^2i} \hspace*{-0.1cm}
\int\hspace*{-0.1cm}\frac{d^4k_2}{\pi^2i}
\hspace*{0.1cm}\frac{\tilde  F_B(k_1,k_2,q)\tilde F_B(k_1,k_2,0)}
{[-k_1v-\bar\Lambda_{q_1q_2}]}
{\rm tr}[\Gamma_2^iS_{q_2}(k_1+k_2)\Gamma_2^fS_{q_1}(k_2-q)\gamma^\mu
S_{q_1}(k_2)]\nonumber
\end{eqnarray}
We have
\begin{eqnarray}
\hspace*{-1cm}& &R^{if; \mu}_{q_1q_2}(v,q)=
\int\limits_0^\infty \hspace*{-0.1cm}ds_1 \tilde F_B^L(6s_1)
\int\limits_0^\infty \hspace*{-0.1cm}ds_2 \tilde F_B^L(6s_2)e^{2s_2q^2}
\int\limits_0^\infty \hspace*{-0.1cm}d^4 \alpha
e^{\alpha_3\bar\Lambda-(\alpha_1+\alpha_4)m_{q_1}^2-\alpha_2m_{q_2}^2}
\nonumber\\
\hspace*{-1cm}&\times&{\rm tr}\biggl[\Gamma_2^i
\biggl(m_{q_2}-\frac{\not\!\partial_1+\not\!\partial_2}{2}\biggr)\Gamma_2^f
\biggl(m_{q_1}-\frac{\not\!\partial_2}{2}-\not\! q\biggr)\gamma^\mu
\biggl(m_{q_1}-\frac{\not\!\partial_2}{2}\biggr)\biggr]\nonumber
\int\hspace*{-0.1cm}\frac{d^4k_1}{\pi^2i} \hspace*{-0.1cm}
\int\hspace*{-0.1cm}\frac{d^4k_2}{\pi^2i} \hspace*{0.1cm}
e^{kAk-2kB}\nonumber
\end{eqnarray}
where the matrices A and B are defined by
\[A_{ij}=\left(
\begin{array}{ll}
 \mbox{$2(s_1+s_2)+\alpha_2$}  &  \hspace*{.5cm}  \mbox{$s_1+s_2+\alpha_2$}\\
 \mbox{$s_1+s_2+\alpha_2$}     &  \hspace*{.5cm}
\mbox{$2(s_1+s_2)+\alpha_1+\alpha_2+\alpha_4$}
\end{array}
\right) \]
\[B_{i}=\left(
\begin{array}{l}
 \mbox{$-s_2q-\alpha_3v/2$} \\
 \mbox{$(s_2+\alpha_1)q$}
\end{array}
\right) \]

The integration over $k_1$ and $k_2$ results in
\begin{eqnarray}
\hspace*{-1cm}& &R^{if; \mu}_{q_1q_2}(v,q)=
\int\limits_0^\infty \hspace*{-0.1cm}ds_1 \tilde F_B^L(6s_1)
\int\limits_0^\infty \hspace*{-0.1cm}ds_2 \tilde F_B^L(6s_2)e^{2s_2q^2}
\int\limits_0^\infty \hspace*{-0.1cm}d^3\alpha
e^{\alpha_3\bar\Lambda-(\alpha_1+\alpha_4)m_{q_1}^2-\alpha_2m_{q_2}^2}
\nonumber\\
\hspace*{-1cm}&\times&{\rm tr}\biggl[\Gamma_2^i
\biggl(m_{q_2}-\frac{\not\!\partial_1+\not\!\partial_2}{2}\biggr)\Gamma_2^f
\biggl(m_{q_1}-\frac{\not\!\partial_2}{2}-\not\! q\biggr)\gamma^\mu
\biggl(m_{q_1}-\frac{\not\!\partial_2}{2}\biggr)\biggr]
\frac{e^{-BA^{-1}B}}{|A|^2}
\nonumber
\end{eqnarray}
Let us, as an example, choose
$\Gamma_2^i=\gamma^\nu$ and $\Gamma_2^f=\gamma^5$.
In the limit $qv=(m_i^2-m_f^2)/(2m_i) \approx 0$ where $m_i$
and $m_f$ are the masses of the initial and the final baryons, respectively,
we find
$$R^{{\rm VP}; \mu\nu}_{q_1q_2}(v,q)=
4i\varepsilon^{\mu\nu\alpha\beta}q^\alpha v^\beta
\cdot\int\limits_0^\infty\hspace*{-0.1cm}
\frac{d^3\alpha\alpha_1\alpha_3}{2|A|^2}
\tilde F_B^2(6z)\{m_{q_1}(A_{11}^{-1}+A_{12}^{-1})-m_{q_2}A_{12}^{-1}\}
$$
\[A_{ij}=\left(
\begin{array}{ll}
 \mbox{$2+\alpha_2$}      &    \hspace*{.2cm}    \mbox{$ 1+\alpha_2$}\\
 \mbox{$ 1+\alpha_2$}     &    \hspace*{.2cm}    \mbox{$2+\alpha_1+\alpha_2$}
\end{array}
\right),
\hspace*{.5cm}
A^{-1}_{ij}=\frac{1}{|A|}\left(
\begin{array}{ll}
 \mbox{$2+\alpha_1+\alpha_2$} &    \hspace*{.2cm}    \mbox{$-(1+\alpha_2)$}\\
 \mbox{$-(1+\alpha_2)$}       &    \hspace*{.2cm}    \mbox{$2+\alpha_2$}
\end{array}
\right) \]
The evaluation of the other remaining matrix elements proceeds along similar
lines.

In the case of one-pion transitions the general expansion of the transition
matrix elements $M_{\rm inv}^\pi(B_Q^i\to B_Q^f\pi)$
into a minimal set of covariants reads
\begin{eqnarray}\label{matrix_el1}
& &\underline{\mbox{One-pion transitions}}\\
M_{inv}^\pi(\Sigma_c\to\Lambda_c\pi) &=&
\frac{1}{\sqrt{3}}g_{\Sigma_c\Lambda_c\pi}
I_1  \bar u(v^\prime) \not\! q \gamma_5 u(v)
\hspace*{.5cm}
\underline{\mbox{p-wave transition}}\nonumber \\
M_{inv}^\pi(\Sigma_c^*\to\Lambda_c\pi) &=&
g_{\Sigma^{*}_c\Lambda_c\pi} I_1
\bar{u}(v^\prime)q_{\mu}u^{\mu}(v)
\hspace*{.5cm}\underline{\mbox{p-wave transition}}\nonumber \\
M_{inv}^\pi(\Lambda_{c1; S}\to\Sigma_c\pi) &=&
f_{\Lambda_{c1; S}\Sigma_c\pi} I_3
\bar{u}(v^{\prime})u(v)
\hspace*{.5cm}\underline{\mbox{s-wave transition}}\nonumber \\
M_{inv}^\pi(\Lambda_{c1; S}^*\to\Sigma_c\pi) &=& \frac{1}{\sqrt{3}}
f_{\Lambda_{c1; S}^{*}\Sigma_c\pi} I_3
\bar{u}(v^{\prime})\gamma_5\not\! q q_{\mu}u^{\mu}(v)
\hspace*{.5cm}
\underline{\mbox{d-wave transition}}
\nonumber
\end{eqnarray}
where the $I_1$ and $I_3$ are the flavor factors
which are directly connected to the flavor coefficients
$C_{\rm flavor}$ (see Eq. (\ref{Pi1})) via the relations
$I_i = f_i \cdot C_{\rm flavor}$, $i=1$ or $3$. The sets of $I_i$ and $f_i$
are given in Table II.
We have also indicated the orbital angular momentum of the pion in
 Eq. (\ref{matrix_el1}).

For the one-photon transitions one similarly has

\begin{eqnarray}\label{matrix_el}
& &\underline{\mbox{One-photon transitions}}\\
M_{inv}^\gamma(\Sigma_c\to\Lambda_c\gamma) &=&
\frac{2i}{\sqrt{3}}f_{\Sigma_c\Lambda_c\gamma}
\bar u(v^\prime) \not\! q \not\!\varepsilon^*(q) u(v)
\hspace*{.5cm}
\underline{\mbox{M1 transition}}\nonumber \\
M_{inv}^\gamma(\Sigma_c^*\to\Lambda_c\gamma) &=&
2f_{\Sigma_c^*\Lambda_c\gamma}
\bar u(v^\prime)\epsilon(\mu \varepsilon^* v q) u^\mu(v)
\hspace*{.5cm}
\underline{\mbox{M1 transition}}\nonumber \\
M_{inv}^\gamma(\Sigma_c^*\to\Sigma_c\gamma) &=&
2if_{\Sigma_c^*\Sigma_c\gamma}
\bar u(v^\prime) \frac{\gamma^\mu \gamma^5}{\sqrt{3}}
u^\nu(v) (q_\mu \varepsilon^*_\nu(q) - q_\nu \varepsilon^*_\mu(q))
\hspace*{.5cm}
\underline{\mbox{M1 transition}}\nonumber \\
M_{inv}^\gamma(\Lambda_{c1; S}\to\Lambda_c\gamma) &=&
\bar u(v^\prime) F_{\Lambda_{c1; S}\Lambda_c\gamma}
[g^{\mu\nu} vq -  v^{\mu}q^{\nu}] \frac{\gamma^\nu\gamma^5}{\sqrt 3}
u(v)\varepsilon^*_\mu(q)
\hspace*{.5cm}
\underline{\mbox{E1 transition}}\nonumber \\
M_{inv}^\gamma(\Lambda_{c1; S}^*\to\Lambda_c\gamma) &=&
\bar u(v^\prime) F_{\Lambda_{c1; S}\Lambda_c\gamma}^*
[g^{\mu\nu} vq -  v^{\mu}q^{\nu}] u^\nu(v)\varepsilon^*_\mu(q)
\hspace*{.5cm}
\underline{\mbox{E1 transition}}\nonumber
\end{eqnarray}
where the couplings are manifestly gauge invariant.
In writing down the one-photon coupling structure in Eq. (\ref{matrix_el})
we have omitted the E2 coupling in the $\Sigma_c^*\to\Lambda_c\gamma$
transition and the M2 coupling in the $\Lambda_{c1; S}^*\to\Lambda_c\gamma$
transitions as predicted by heavy quark symmetry \cite{KKP}. Since heavy quark
symmetry is manifest in our model calculation the coupling structure
(\ref{matrix_el}) is sufficient for our purposes.
In the heavy quark limit the coupling constants in
Eqs. (\ref{matrix_el1}) and (\ref{matrix_el}) become flavor independent.
Also some of the above coupling constants become related in the heavy quark
limit. The relations read \cite{KKP,Yan,Cho}
\begin{eqnarray}\label{rel_con}
g_{\Sigma_c\Lambda_c\pi}=g_{\Sigma^{*}_c\Lambda_c\pi}=g,
\hspace*{1cm}
f_{\Sigma_c\Lambda_c\gamma}=f_{\Sigma^*_c\Lambda_c\gamma}=f,
\hspace*{1cm}
F_{\Lambda_{c1; S}\Lambda_c\gamma}=F_{\Lambda^{*}_{c1; S}\Lambda_c\gamma}=F
\end{eqnarray}

Returning to our model calculation the coupling constant $f$ can be
represented as
\begin{eqnarray}
f&=&(\mu_1-\mu_2)
\frac{R_{\Sigma_Q\Lambda_Q\gamma}}{\sqrt{R_{\Lambda_Q}}\sqrt{R_{\Sigma_Q}}}
\\[2mm]
R_{\Sigma_Q\Lambda_Q\gamma}&=&\frac{1}{4}
\int\limits_0^\infty\hspace*{-.1cm}d^3\alpha\alpha_3(\alpha_1+\alpha_2)
\tilde F^2_B(6z)\frac{A_{11}^{-1}}{|A|^2}
\nonumber\\[2mm]
R_{B_Q}&=&
\int\limits_0^\infty\hspace*{-.1cm}d^3\alpha\alpha_3
\frac{\tilde F^2_B(6z)}{|A|^2}
\biggl\{1+d_{B_Q}\frac{\alpha_3}{m^2_q}\frac{\partial z}{\partial\alpha_3}
-\frac{\alpha_3^2}{4m^2_q}A_{12}^{-1}(A_{11}^{-1}+A_{12}^{-1})\biggr\}
\nonumber
\end{eqnarray}
where $\mu_i=e_i/(2m_q)$ is the magnetic moment of the i-th light quark.
Here
$$
z=\frac{\alpha_3^2}{4}A_{11}^{-1}+m^2_q(\alpha_1+\alpha_2)
-\bar\Lambda\alpha_3,
\hspace*{1cm}
d_{B_Q}=\left\{
\begin{array}{ll}
1 & \,\,\, \mbox{for} \,\,\, B_Q=\Lambda_{Q}\\
\frac{1}{2} & \,\,\, \mbox{for} \,\,\, B_Q=\Sigma_{Q}\\
\end{array}
\right.
$$
The calculation of the other coupling factors proceeds along similar lines.
In addition to the relations (\ref{rel_con}) there is the identity
between $f_{\Sigma_c^*\Lambda_c\gamma}$ and $f_{\Sigma_c^*\Sigma_c\gamma}$
couplings obtained in the constituent quark model \cite{HKLT,HKT}:
$$f_{\Sigma_c^*\Lambda_c\gamma}/f_{\Sigma_c^*\Sigma_c\gamma} =
(\mu_1-\mu_2)/(\mu_1+\mu_2)=3$$
In our model the light diquark current in $\Sigma_c$-baryon
is different than that in $\Lambda_c$-baryon (see TABLE I).
However numerically the latter relation are reproduced with an
accuracy 1$\%$.

One can then go on and calculate the one-pion and one-photon decay rates
using the general formula
\begin{eqnarray}\label{rate1}
\Gamma (B_Q^i\to B_Q^fX)
= \frac{1}{2J+1} \quad \frac{ \mid \vec{q} \mid}{8 \pi
M_{B_Q}^{2}}\sum_{spins} \mid M^{\pi}_{inv} (B_Q^i\to B_Q^fX) \mid^{2}
\end{eqnarray}
where $\mid \vec{q} \mid $ is the pion (photon) momentum in the rest frame
of the decaying baryon. In terms of the coupling constants (\ref{rel_con})
one obtains
\begin{eqnarray}\label{all_rates}
\Gamma\left( \Sigma_c \rightarrow \Lambda_c \pi \right)&=& g^2I_1^2
\frac{{\mid \vec{q} \mid}^3}{6\pi} \frac{M_{\Lambda_c}}{M_{\Sigma_c}}
\nonumber\\
\Gamma\left( \Sigma^*_c \rightarrow \Lambda_c \pi \right)&=& g^2I_1^2
\frac{{\mid\vec{q}\mid}^3}{6\pi} \frac{M_{\Lambda_c}}{M_{\Sigma^*_c}}
\nonumber\\
\Gamma\left(\Lambda_{c1; S}\rightarrow\Sigma_c\pi \right)
&=&f^2_{\Lambda_{c1; S}\Sigma\pi} I_3^2\frac{\mid \vec{q} \mid}{2\pi}
\frac{M_{\Sigma_c}}{M_{\Lambda_{c1; S}}}\nonumber\\
\Gamma\left(\Lambda^{*}_{c1; S}\rightarrow\Sigma_c\pi\right)
&=& f^2_{\Lambda^{*}_{c1; S}\Sigma\pi} I_3^2
\frac{{\mid\vec{q}\mid}^5}{18\pi}\frac{M_{\Sigma_c}}{M_{\Lambda^*_{c1; S}}}
\\
\Gamma\left(\Sigma_c\rightarrow \Lambda_c \gamma \right)&=&
\frac{4}{3\pi} f^2 {\mid \vec{q} \mid}^3 \frac{M_{\Lambda_c}}{M_{\Sigma_c}}
\nonumber\\
\Gamma\left(\Sigma^*_c\rightarrow \Lambda_c\gamma\right)&=&
\frac{4}{3\pi} f^2 {\mid \vec{q} \mid}^3 \frac{M_{\Lambda_c}}{M_{\Sigma_c^*}}
\nonumber\\
\Gamma\left(\Sigma^*_c\rightarrow \Sigma_c\gamma\right)&=&
\frac{4}{9\pi} f^2_{\Sigma_c^*\Sigma_c\gamma}
{\mid \vec{q} \mid}^3 \frac{M_{\Sigma_c}}{M_{\Sigma_c^*}}
\nonumber\\
\Gamma\left(\Lambda_{c1; S}\rightarrow\Lambda_c\gamma \right)&=&
\frac{1}{3\pi} F^2 {\mid \vec{q} \mid}^3
\frac{M_{\Lambda_c}}{M_{\Lambda_{c1; S}}}
\nonumber\\
\Gamma\left(\Lambda_{c1; S}^*\rightarrow\Lambda_c\gamma \right)&=&
\frac{1}{3\pi} F^2 {\mid \vec{q} \mid}^3
\frac{M_{\Lambda_c}}{M_{\Lambda_{c1; S}^*}}
\nonumber
\end{eqnarray}
The unknown masses of the excited bottom baryons
$\Lambda_{b1; S}$ and $\Lambda_{b1; S}^*$
are estimated from the heuristic relation:
$m_{\Lambda_{b1; S}^{(*)}} = m_{\Lambda_{c1; S}^{(*)}} +
(m_{\Lambda_{b}^{0}} - m_{\Lambda_{c}^{+}})$.

\section{Results}

We now present our numerical results for
the strong and radiative transitions of heavy baryons.
In Table III we list our results for the one-pion coupling constants.
For comparison we also give the results of the Light-Front (LF) Quark Model
\cite{TOK} where the masses of light quarks are varied in the range
$m_{u(d)}=280\pm 60$ MeV and $m_s=m_{u(d)}+150$ MeV $=430\pm 60$ MeV.
Our predictions for the one-pion coupling constants
are in qualitative agreement with the Light-Front quark model prediction.
As in the Light-Front
quark model \cite{TOK} the strength of single pion couplings of the
charmed baryon ground-state to the antisymmetric P wave states are
suppressed with respect to those of the symmetric multiplet.
The coupling values calculated in the Light-Front approach show a
strong dependence on the masses of the light quarks.
The results of Ref. \cite{TOK} are quite close to our results for
the mass choice $m_{u(d)}=220$ MeV and $m_s=370$ MeV. However, such low values
of the constituent quark masses are excluded from the analysis of the magnetic
moments and charge radii of nucleons.
The smaller value of $f_{\Lambda_{c1; S}^{*}\Sigma_c\pi}$
is welcome when one compares the results for exclusive one-pion rates of the
$\Lambda_{c1; S}^{*}$ (see, Table IV) with experimental
data \cite{PDG}. It is seen that our predictions are consistent with current
experimental estimates whereas the Light-Front model results lie above
the experimental rates. Also our predictions for $d$-wave
transitions can be seen to be consistent with the bound on the total rate of
the $\Lambda_{c1; S}^*$ by summing up the three exclusive one-pion
rates of the $\Lambda_{c1; S}^{*}$ and comparing the sums
to the total experimental rate $\Gamma({\Lambda^{*}_{c1; S}}) <1.9$ MeV. 
From the results of our model
calculation we obtain $\Gamma(\Lambda_{c1; S}^{*}) > 0.25 \pm 0.03 $ MeV
consistent with the experimental results.
All our results for the one-pion decay rates
of charm baryons are collected in Table IV.
The uncertainties for the calculated rates
reflect the experimental errors in the charm baryon masses (see, Table I). For
comparison we have also listed the predictions of the Light-Front quark
model \cite{TOK} and experimental results, where available.
More precise data on the one-pion transitions of the
excited $\Lambda_{c1; S}$ baryon states to the ground states will allow for
a more detailed comparison with the model
predictions of dynamical models such as described in our approach and
in the Light-Front quark model.

A few comments should be done concerning the relations of RTQM results to
other approaches. The first comment concerns the baryon one-pion
coupling constants appearing in Eq. (18). The chiral formalism employed in
\cite{Yan} implies that all such constants are proportional to the factor
$1/f_\pi$ associated with the pion field. In RTQM approach they are described
by Eq.(9) and are proportional to the pion-quark coupling $g_\pi$ which is
determined by the compositeness condition discussed in Sec.II.
However, it appears that the Goldberger-Treiman relation $g_\pi=2m_q/f_\pi$ is
valid with an accuracy of a few percent. Hence the pion constant $f_\pi$
effectively appears as a dimensional parameter in our expressions for
the baryon one-pion coupling constants.

The second comment concerns the relation of the RTQM approach to 
the constituent quark model based on the $SU(4)\otimes O(3)$ symmetry for 
the light diquark system \cite{HKLT,HKT}.
Exploiting this symmetry one can reduce the number of the effective coupling
constants both for one-pion and one-photon transitions \cite{HKLT,HKT}.
For example, the ratio of the two coupling constants  
$f_{\Lambda_{c1; S}\Sigma_c\pi}$ and $f_{\Sigma_{c0; S}\Lambda_c\pi}$ 
that govern the $s$-wave transitions
$\Lambda_{c1; S}\to\Sigma_c\pi$ and $\Sigma_{c0; S}\to\Lambda_c\pi$ is 
predicted to be 
$f_{\Lambda_{c1; S}\Sigma_c\pi}/f_{\Sigma_{c0; S}\Lambda_c\pi}=-1/\sqrt{3}$ 
in the $SU(4)\otimes O(3)$ model. We emphasize that 
the $SU(4)\otimes O(3)$ symmetry is not realized in the RTQM approach. 
For instance, differing from the $SU(4)\otimes O(3)$ approach the light 
diquark current in $\Sigma_{c0; S}$-baryon
is not related to that in $\Lambda_{c1; S}$ (see TABLE I). 
Yet, numerically the $SU(4)\otimes O(3)$ relations are reproduced with an 
amazing accuracy of around 1$\%$. For the two above processes 
$\Lambda_{c1; S}\to\Sigma_c\pi$ and $\Sigma_{c0; S}\to\Lambda_c\pi$ we find 
numerically 
$f_{\Lambda_{c1; S}\Sigma_c\pi}/f_{\Sigma_{c0; S}\Lambda_c\pi}=-0.58$ almost 
identical to the ratio $-1/\sqrt{3}$ in the $SU(4)\otimes O(3)$ approach. 
Note that the $SU(4)\otimes O(3)$ symmetry can be implemented explicitly
in the RTQM by replacing the light quark propagator by a constant value and
by modifying the spinor structure of the  baryon-quark vertices
as was demonstrated in \cite{IKLR1}.

Numerical results for the one-photon decay rates are listed in Table V.
As for one-pion rates the errors in our rate values reflect the experimental
errors in the charm baryon masses \cite{PDG,CLEO4} (see Table I).
For the sake of comparison we also list the results of the model
calculations \cite{Cho}-\cite{Chow} mentioned earlier on. Our results are
quite close to the results of the nonrelativistic quark model \cite{Cheng1}.
In \cite{Cho} the coupling strengths were parameterized in terms of the
unknown effective coupling parameters $c_{RS}$ and $c_{RT}$. 
A first rough estimate of the
unknown coupling parameters can be obtained by setting them equal to 1 on
dimensional grounds \cite{Cho}. As is evident from Table V such an estimate
is basically supported by our dynamical calculation. We do not agree with
the predictions on the charm and bottom p-wave decay rates of \cite{Chow}
except for the $\Lambda^*_{b1; S}\to \Lambda_b^0\gamma$ rate where we are
closer to the rate calculated in \cite{Chow}.

Recently the radiative decays of bottom baryons were studied with the use
of the light-cone QCD sum rules \cite{ZD} in the leading order of heavy
quark effective theory. For the decay rates of the $\Sigma_b$ and
$\Sigma_b^*$ baryons to $\Lambda_b^0\gamma$ the authors of \cite{ZD}
obtained
$$
\Gamma(\Sigma_b\to\Lambda_b\gamma)=\alpha_{\rm eff}{\mid \vec{q} \mid}^3
\,\,\,\,\, \mbox{and}\,\,\,\,\,
\Gamma(\Sigma_b^*\to\Lambda_b\gamma)=\alpha^*_{\rm eff}{\mid \vec{q}\mid}^3
$$
where the couplings $\alpha_{\rm eff}$ and $\alpha^*_{\rm eff}$ are
approximately  equal to each other. The authors of \cite{ZD} quote
$\alpha_{\rm eff}\approx\alpha^*_{\rm eff}\approx 0.03 $ GeV~$^{-2}$.
In order to compare our model results with the results in \cite{ZD} we set
$M_{\Lambda_Q}=M_{\Sigma_Q}$ in Eq. (\ref{all_rates}). We then obtain
$\alpha_{\rm eff} = 4f^2/(3\pi) \approx 0.015 $GeV$^{-2}$ which is one-half of
the prediction of Ref.~\cite{ZD}.

One has to remark that the rate of the $\Sigma_c^*\to\Sigma_c\gamma$ decay is
suppressed in comparison with the rate of
$\Sigma_c^*\to\Lambda_c\gamma$ decay as
\begin{eqnarray}
\frac{\Gamma\left(\Sigma^*_c\rightarrow \Sigma_c\gamma\right)}
     {\Gamma\left(\Sigma^*_c\rightarrow \Lambda_c\gamma\right)}
= \frac{1}{3}
\biggl(\frac{f_{\Sigma_c^*\Sigma_c\gamma}}{f_{\Sigma_c^*\Lambda_c\gamma}}
\biggr)^2 \frac{M_{\Sigma_c}}{M_{\Lambda_c}}
\biggl(\frac{M^2_{\Sigma_c^*}-M^2_{\Sigma_c}}
{M^2_{\Sigma_c^*}-M^2_{\Lambda_c}}\biggr)^3 \approx 10^{-3}
\end{eqnarray}
This estimate coincides with the prediction of the constituent quark
model \cite{HKLT,HKT}.

\section{Conclusion}

Using the same Relativistic Three-Quark Model (RTQM) employed before
successfully in  studies of semileptonic decays of heavy baryons \cite{ILKK},
we have analyzed strong (one-pion) and radiative decays of heavy baryons.
We have obtained predictions for the values of couplings of charmed
baryons with pions and for the rates of the one-pion transitions
$B^i_c(p)\to B^f_c(p^\prime)+\pi(q)$. We have compared our results with
those obtained in Light-Front Quark-Model \cite{TOK}.
We found that as in the Light-Front
quark model \cite{TOK} the strength of single pion couplings of the
charmed baryon ground-state to the antisymmetric P wave states are
suppressed with respect to those of the symmetric multiplet.

We have also obtained predictions for the rates of the one-photon transitions
$B^i_Q(p)\to B^f_Q(p^\prime)+\gamma(q)$. We have compared our results with
the results of other model calculations~\cite{Cho}-\cite{Chow}. Unfortunately,
there is no data on the one-photon transitions yet to compare our results
with. For the one-photon decays
from the $p$-wave states $\Lambda_{c1; S}\to\Lambda_c+\gamma$ and
$\Lambda_{c1; S}^*\to\Lambda_c+\gamma$ our predicted rates are one order of
magnitude below the upper limits given by the experiments calling for an
one-order of magnitude improvement on the experimental upper limits.
Although the $\Xi_c^\prime \to\Xi_c + \gamma$ one-photon decays have now been
seen \cite{CLEO4} it will be close to impossible to obtain rate values for
these decays because the $\Xi_c^\prime$-states are far too narrow to be
experimentally resolvable. The total
widths of the $\Sigma_c$, $\Sigma_c^*$ and $\Xi_c^*$ states are larger
because of their strong decays via one-pion emission. In fact the widths of
the $\Sigma_c^{*++}$ and $\Sigma_c^{*0}$ have been determined \cite{PDG}. One
can hope that one-pion branching ratios can be experimentally determined for
the $\Sigma_c$, $\Sigma_c^*$ and $\Xi_c^*$ one-photon decay modes in the
near future. We are looking forward to compare the predictions of
the Relativistic Three-Quark Model for the calculated rates with
future experimental data.

\section*{Acknowledgments}

\noindent
M.A.I, V.E.L and A.G.R thank Mainz University for hospitality
where a part of this work was completed. This work was supported
in part by the Heisenberg-Landau Program
and by the BMBF (Germany) under contract 06MZ865. J.G.K. acknowledges
partial support by the BMBF (Germany) under contract 06MZ865.

\appendix
\section{Gauge invariance in RTQM}
In this Appendix we provide a detailed discussion of how
gauge invariance is maintained in the one-photon transitions
of excited charm baryon states. As an example we consider the decay
$\Lambda_{c1; S}^*\to\Lambda_c\gamma$. To begin with we keep the
heavy quark mass and the heavy baryon masses finite.
The corresponding matrix element has the form
\begin{eqnarray}
M_{inv}^\gamma(\Lambda_{c1; S}^*\to\Lambda_c\gamma) =
\bar u(p^\prime) \Lambda^{\mu\nu}(p,p^\prime) u_\nu(p)\varepsilon^*_\mu(q)
\end{eqnarray}
Here the vertex function $\Lambda^{\mu\nu}(p,p^\prime)$
is given by the expression
\begin{eqnarray}
\Lambda^{\mu\nu}(p,p^\prime) = \Lambda^{\mu\nu}_{\Delta, Q}(p,p^\prime) +
\Lambda^{\mu\nu}_{\Delta, q}(p,p^\prime) +
\Lambda^{\mu\nu}_{\rm lcd}(p,p^\prime) +
\Lambda^{\mu\nu}_{\rm ncd}(p,p^\prime) +
\Lambda^{\mu\nu}_{\rm pole}(p,p^\prime)
\end{eqnarray}
Here
$\Lambda^{\mu\nu}_{\Delta, Q}(p,p^\prime)$,
$\Lambda^{\mu\nu}_{\Delta, q}(p,p^\prime)$,
$\Lambda^{\mu\nu}_{\rm lcd}(p,p^\prime)$,
$\Lambda^{\mu\nu}_{\rm ncd}(p,p^\prime)$
and $\Lambda^{\mu\nu}_{\rm pole}(p,p^\prime)$
are the partial contributions coming from
the triangle diagram with the photon emitted by the heavy quark line,
the triangle diagrams with the photon emitted by the light quark lines,
the "local contact" diagrams (lcd),
the "nonlocal contact" diagrams (ncd) and the pole diagrams:
their functional form is given by
\begin{eqnarray}
\Lambda^{\mu\nu}_ {\Delta, Q}(p,p^\prime)&=& e_Qg_{eff}
\int\hspace*{-0.1cm}\frac{d^4k_1}{\pi^2i} \hspace*{-0.1cm}
\int\hspace*{-0.1cm}\frac{d^4k_2}{4\pi^2i} \hspace*{0.1cm}
\tilde F_B^2(k_1,k_2,0)S_Q(k_1+p^\prime)\gamma^\mu S_Q(k_1+p) k_1^\nu
\nonumber\\
&\times&{\rm tr}[\gamma_5 S_q(k_1+k_2) \gamma_5 S_q(k_2)]\nonumber\\
\Lambda^{\mu\nu}_ {\Delta, q}(p,p^\prime)&=& e_qg_{eff}
\int\hspace*{-0.1cm}\frac{d^4k_1}{\pi^2i} \hspace*{-0.1cm}
\int\hspace*{-0.1cm}\frac{d^4k_2}{4\pi^2i} \hspace*{0.1cm}
\tilde F_B(k_1,k_2,0)\tilde F_B(k_1,k_2,q) S_Q(k_1+p) k_1^\nu \nonumber\\
&\times&{\rm tr}[\gamma_5 S_q(k_1+k_2) \gamma_5 S_q(k_2-q)\gamma^\mu S_q(k_2)]
\nonumber\\
\Lambda^{\mu\nu}_ {\rm lcd}(p,p^\prime)&=& -g^{\mu\nu}e_qg_{eff}
\int\hspace*{-0.1cm}\frac{d^4k_1}{\pi^2i} \hspace*{-0.1cm}
\int\hspace*{-0.1cm}\frac{d^4k_2}{4\pi^2i} \hspace*{0.1cm}
\tilde F_B(k_1,k_2,0)\tilde F_B(k_1,k_2,-q) S_Q(k_1+p^\prime)
\nonumber\\
&\times&{\rm tr}[\gamma_5 S_q(k_1+k_2) \gamma_5 S_q(k_2)]
\nonumber\\
\Lambda^{\mu\nu}_ {\rm ncd}(p,p^\prime)&=& e_qg_{eff}
\int\hspace*{-0.1cm}\frac{d^4k_1}{\pi^2i} \hspace*{-0.1cm}
\int\hspace*{-0.1cm}\frac{d^4k_2}{4\pi^2i} \hspace*{0.1cm}
\tilde F_B(k_1,k_2,0) k_1^\nu
{\rm tr}[\gamma_5 S_q(k_1+k_2) \gamma_5 S_q(k_2)]\nonumber\\
&\times&[S_Q(k_1+p)(k_1-k_2+q)^\mu
\frac{\tilde F_B(k_1,k_2,0)-\tilde F_B(k_1,k_2,q)}
{(k_1-k_2)q+q^2}\nonumber\\
&-&S_Q(k_1+p^\prime)(k_2-k_1+q)^\mu
\frac{\tilde F_B(k_1,k_2,0) - \tilde F_B(k_1,k_2,-q)}
{(k_2-k_1)q+q^2}] \nonumber\\
\Lambda^{\mu\nu}_ {\rm pole}(p,p^\prime)&=& (e_Q+e_q)g_{eff}
\int\hspace*{-0.1cm}\frac{d^4k_1}{\pi^2i} \hspace*{-0.1cm}
\int\hspace*{-0.1cm}\frac{d^4k_2}{4\pi^2i} \hspace*{0.1cm}
\tilde F_B^2(k_1,k_2,0)k_1^\nu
{\rm tr}[\gamma_5 S_q(k_1+k_2) \gamma_5 S_q(k_2)]\nonumber\\
&\times&\biggl\{ \gamma^\mu\frac{1}{m_f-\not\! p}S_Q(k_1+p) +
S_Q(k_1+p^\prime)\frac{1}{m_i-\not\! p^\prime}\gamma^\mu\biggr\}
\nonumber
\end{eqnarray}
where $m_i$ and $m_f$ are the masses of the initial and the final baryons,
respectively; $e_Q=2e/3$ is the charge of the $c$-quark,
$e_q=e_{q_1}+e_{q_2}$ is the sum of charges of the light quarks in the
heavy baryon ($e_q=e/3$ in the present case).

It is a straightforward exercise to proof that the matrix element
$M_{inv}^\gamma(\Lambda_{c1; S}^*\to\Lambda_c\gamma)$ is gauge invariant, i.e.
$q_\mu \cdot \bar u(p^\prime) \Lambda^{\mu\nu}(p,p^\prime) u_\nu(p) = 0$ when
both initial and final baryons are on their mass-shell:
$\not p u(p) = m_i u(p)$ and
$\bar u(p^\prime)\not p^\prime = \bar u(p^\prime ) m_f$.
We make use of the well-known Ward-Takahashi identities
$$\not \! q = S^{-1}_Q(k+p^\prime) - S^{-1}_Q(k+p),
\hspace*{1cm} \not \! q = S^{-1}_q(k-q) - S^{-1}_q(k)$$
and obtain
\begin{eqnarray}
q_\mu \cdot \bar u(p^\prime)
\Lambda^{\mu\nu}_ {\Delta, Q}(p,p^\prime)u_\nu(p)
&=& e_Qg_{eff}
\int\hspace*{-0.1cm}\frac{d^4k_1}{\pi^2i} \hspace*{-0.1cm}
\int\hspace*{-0.1cm}\frac{d^4k_2}{4\pi^2i} \hspace*{0.1cm}
\tilde F_B^2(k_1,k_2,0) k_1^\nu
{\rm tr}[\gamma_5 S_q(k_1+k_2) \gamma_5 S_q(k_2)]
\nonumber\\
&\times&
\bar u(p^\prime)[S_Q(k_1+p)-S_Q(k_1+p^\prime)]u_\nu(p)
\nonumber\\
q_\mu \cdot  \bar u(p^\prime)\Lambda^{\mu\nu}_ {\Delta, q}(p,p^\prime)u_\nu(p)
&=& e_qg_{eff}
\int\hspace*{-0.1cm}\frac{d^4k_1}{\pi^2i} \hspace*{-0.1cm}
\int\hspace*{-0.1cm}\frac{d^4k_2}{4\pi^2i} \hspace*{0.1cm}
\tilde F_B(k_1,k_2,0) k_1^\nu
{\rm tr}[\gamma_5 S_q(k_1+k_2) \gamma_5 S_q(k_2)]
\nonumber\\
&\times&\bar u(p^\prime)[\tilde F_B(k_1,k_2,q) S_Q(k_1+p)
-\tilde F_B(k_1,k_2,-q) S_Q(k_1+p^\prime)]u_\nu(p)
\nonumber\\
&+&q^{\nu}e_qg_{eff}
\int\hspace*{-0.1cm}\frac{d^4k_1}{\pi^2i} \hspace*{-0.1cm}
\int\hspace*{-0.1cm}\frac{d^4k_2}{4\pi^2i} \hspace*{0.1cm}
\tilde F_B(k_1,k_2,0)\tilde F_B(k_1,k_2,-q)\nonumber\\
&\times&{\rm tr}[\gamma_5 S_q(k_1+k_2) \gamma_5 S_q(k_2)]
\bar u(p^\prime)
S_Q(k_1+p^\prime)u_\nu(p)
\nonumber\\
q_\mu \cdot \bar u(p^\prime)\Lambda^{\mu\nu}_ {\rm lcd}(p,p^\prime)u_\nu(p)
&=& -q^{\nu}e_qg_{eff}
\int\hspace*{-0.1cm}\frac{d^4k_1}{\pi^2i} \hspace*{-0.1cm}
\int\hspace*{-0.1cm}\frac{d^4k_2}{4\pi^2i} \hspace*{0.1cm}
\tilde F_B(k_1,k_2,0)\tilde F_B(k_1,k_2,-q)
\nonumber\\
&\times&{\rm tr}[\gamma_5 S_q(k_1+k_2) \gamma_5 S_q(k_2)]
\bar u(p^\prime)S_Q(k_1+p^\prime) u_\nu(p)
\nonumber\\
q_\mu \cdot \bar u(p^\prime) \Lambda^{\mu\nu}_ {\rm ncd}(p,p^\prime) u(p)
&=& e_qg_{eff} \int\hspace*{-0.1cm}\frac{d^4k_1}{\pi^2i} \hspace*{-0.1cm}
\int\hspace*{-0.1cm}\frac{d^4k_2}{4\pi^2i} \hspace*{0.1cm}
\tilde F_B(k_1,k_2,0) k_1^\nu
{\rm tr}[\gamma_5 S_q(k_1+k_2) \gamma_5 S_q(k_2)]\nonumber\\
&\times&\biggl\{
[\tilde F_B(k_1,k_2,0) -
\tilde F_B(k_1,k_2,q)]
\bar u(p^\prime)S_Q(k_1+p)u_\nu(p)\biggr\}\nonumber\\
&-& [\tilde F_B(k_1,k_2,0)-\tilde F_B(k_1,k_2,-q)]
\bar u(p^\prime)S_Q(k_1+p^\prime)u_\nu(p)\biggr\}
\nonumber\\
q_\mu \cdot \bar u(p^\prime)\Lambda^{\mu\nu}_ {\rm pole}(p,p^\prime)u_\nu(p)
&=& (e_Q+e_q)g_{eff}
\int\hspace*{-0.1cm}\frac{d^4k_1}{\pi^2i} \hspace*{-0.1cm}
\int\hspace*{-0.1cm}\frac{d^4k_2}{4\pi^2i} \hspace*{0.1cm}
\tilde F_B^2(k_1,k_2,0)k_1^\nu\nonumber\\
&\times&{\rm tr}[\gamma_5 S_q(k_1+k_2) \gamma_5 S_q(k_2)]
\bar u(p^\prime)[S_Q(k_1+p^\prime) - S_Q(k_1+p)] u_\nu(p)
\nonumber
\end{eqnarray}
Summing up the five contributions we arrive at
$q_\mu \cdot \bar u(p^\prime) \Lambda^{\mu\nu}(p,p^\prime) u_\nu(p) = 0$.

In the heavy quark limit, when the masses of the charm quark and the heavy
baryons go to infinity, the triangle diagram where the photon is emitted by
the heavy quark $(\Lambda^{\mu\nu}_ {\Delta, Q}(p,p^\prime))$
and the pole diagrams $(\Lambda^{\mu\nu}_ {{\rm pole}, Q}(p,p^\prime))$
vanish as $1/M$. The sum of the remaining contributions
$\Lambda^{\mu\nu}_ {\Delta, q}(p,p^\prime)$,
$\Lambda^{\mu\nu}_ {{\rm lcd}, Q}(p,p^\prime)$ and
$\Lambda^{\mu\nu}_ {{\rm ncd}, Q}(p,p^\prime)$ can be seen to have
the following gauge-invariant Lorentz structure also appearing in Eq.(19)
$$\Lambda^{\mu\nu}(v,q) = F (g^{\mu\nu}vq - v^\mu q^\nu)$$
where $F$ is an effective coupling constant of the charm baryons with the
photon.

\newpage

%
\begin{table}[t]
\caption{
Quantum numbers of heavy baryons.
($\lambda_u=\rm diag(1,0,0), \lambda_d=\rm diag(0,1,0), $\\
$\stackrel{\leftrightarrow}\partial_\mu^\pm =
\stackrel{\leftarrow}\partial_\mu \pm
\stackrel{\rightarrow}\partial_\mu$).
}
\begin{tabular}{|c|c|c|c|c|c|}
Baryon&$\;J^P\;$&Quark & $\Gamma_1\otimes C\Gamma_2$
& $\; \lambda_{B_Q}\;$& $\;$Mass (MeV) \cite{PDG} $\;$\\
      & & Content & & &\\
\hline
$\Lambda_c^+$    & ${\frac{1}{2}}^+$&c[ud] &$I\otimes C\gamma^5$ & $i\lambda_2/2$ & $2284.9\pm 0.6$\\
\hline
$\Xi_c^+$        & ${\frac{1}{2}}^+$&c[us] &$I\otimes C\gamma^5$ & $i\lambda_5/2$ & $2465.6\pm 1.4$\\
\hline
$\Xi_c^0$        & ${\frac{1}{2}}^+$&c[ds] &$I\otimes C\gamma^5$ & $i\lambda_7/2$ & $2470.3\pm 1.8$\\
\hline
$\Xi_c^{+\prime}$& ${\frac{1}{2}}^+$&c\{us\} &$\gamma^\mu\gamma^5 \otimes C\gamma_\mu$ & $\lambda_4/(2\sqrt{3})$ & $2573.4\pm 3.1$\\
\hline
$\Xi_c^{0\prime}$& ${\frac{1}{2}}^+$&c\{ds\} &$\gamma^\mu\gamma^5 \otimes C\gamma_\mu$ & $\lambda_6/(2\sqrt{3})$ & $2577.3\pm 3.2$\\
\hline
$\Sigma_c^{++}$  & ${\frac{1}{2}}^+$&c\{uu\} &$\gamma^\mu\gamma^5 \otimes C\gamma_\mu$ & $\lambda_u/\sqrt{6}$ & $2452.8\pm 0.6$\\
\hline
$\Sigma_c^+$     & ${\frac{1}{2}}^+$&c\{ud\} &$\gamma^\mu\gamma^5 \otimes C\gamma_\mu$ & $\lambda_1/(2\sqrt{3})$ & $2453.6\pm 0.9$\\
\hline
$\Sigma_c^0$     & ${\frac{1}{2}}^+$&c\{dd\} &$\gamma^\mu\gamma^5 \otimes C\gamma_\mu$ & $\lambda_d/\sqrt{6}$ & $2452.2\pm 0.6$\\
\hline
$\Xi_c^{*+}$     & ${\frac{3}{2}}^+$&c\{us\} &$I \otimes C\gamma_\mu$ & $\lambda_4/2$ & $2644.6\pm 2.1$\\
\hline
$\Xi_c^{*0}$     & ${\frac{3}{2}}^+$&c\{ds\} &$I \otimes C\gamma_\mu$ & $\lambda_6/2$ & $2643.8\pm 1.8$\\
\hline
$\Sigma_c^{*++}$ & ${\frac{3}{2}}^+$&c\{uu\} &$I \otimes C\gamma_\mu$ & $\lambda_u/\sqrt{2}$ & $2519.4\pm 1.5$\\
\hline
$\Sigma_c^{*0}$    & ${\frac{3}{2}}^+$&c\{dd\} &$I \otimes C\gamma_\mu$ & $\lambda_d/\sqrt{2}$ & $2517.5\pm 1.4$\\
\hline
$\Lambda_{c1; S}$   & ${\frac{1}{2}}^-$  &c[ud]
&$\gamma^\mu\gamma^5\otimes C\gamma^5\stackrel{\leftrightarrow}\partial_\mu^+$
& $i\lambda_2/(2\sqrt{3})$ & $2593.9\pm 0.8$\\
\hline
$\Sigma_{c0; S}$   & ${\frac{1}{2}}^-$  &c\{ud\}
&$I \otimes C\gamma^\mu \stackrel{\leftrightarrow}\partial_\mu^+$
& $\lambda_1/(2\sqrt{3})$ & $2670$\\
\hline
$\Sigma_{c1; A}$   & ${\frac{1}{2}}^-$  &c[ud]
&$\gamma^\mu\gamma^5\otimes C\gamma^5\stackrel{\leftrightarrow}\partial_\mu^-$
& $i\lambda_2/(2\sqrt{3})$ & $2670$\\
\hline
$\Lambda_{c1; S}^*$ & ${\frac{3}{2}}^-$  &c[ud]  &
$I \otimes C\gamma^5\stackrel{\leftrightarrow}\partial_\mu^+ $
& $i\lambda_2/2$ & $2626.6\pm 0.8$\\
\hline
$\Xi_{c1; S}^*$     & ${\frac{3}{2}}^-$  &c[us]
&$I \otimes C\gamma^5 \stackrel{\leftrightarrow}\partial_\mu^+ $
& $i\lambda_5/2$ & $2815.0 \pm 2.1$ \\
\hline
$\Sigma_{c1; A}^*$  & ${\frac{3}{2}}^-$  &c[ud]
&$I\otimes C\gamma^5\stackrel{\leftrightarrow}\partial_\mu^-$
& $i\lambda_2/(2\sqrt{3})$ & $2701$\\
\hline
$\Lambda_b$    & ${\frac{1}{2}}^+$&b[ud] &$I\otimes C\gamma^5$
& $i\lambda_2/2$ & $5624\pm 9$\\
\hline
$\Lambda_{b1; S}$    & ${\frac{1}{2}}^+$&b[ud]
&$\gamma^\mu\gamma^5\otimes C\gamma^5\stackrel{\leftrightarrow}\partial_\mu^+$
& $i\lambda_2/(2\sqrt{3})$ & $5933\pm 10$\\
\hline
$\Lambda_{b1; S}^*$  & ${\frac{1}{2}}^+$&b[ud]
&$I \otimes C\gamma^5 \stackrel{\leftrightarrow}\partial_\mu^+$
& $i\lambda_2/2$ & $5966\pm 10$\\
\end{tabular}
\end{table}

\begin{table}[t]
\caption{
Flavor coefficients $I_1$, $I_3$ and $f_1$, $f_3$.
}
 \begin{tabular}{|c|c|c||c|c|c|}
 Decay mode & $I_1$  &  $f_1$ & Decay mode & $I_3$ & $f_3$ \\
 \hline\hline
$\Sigma^{+}_{c} \rightarrow \Lambda_c\pi^{0} $  & $1$ & $\sqrt{3}/2$
&$\Lambda_{c1}(2593) \rightarrow \Sigma^{0}_c\pi^{+} $ & $1$ & $3/2$ \\
\hline
$\Sigma^{0}_{c} \rightarrow \Lambda_c\pi^{-} $  & $1$ & $\sqrt{3}/2$
&$\Lambda_{c1}(2593) \rightarrow \Sigma^{+}_c\pi^{0} $ & $1$ & $3/2$ \\
\hline
$\Sigma^{++}_{c} \rightarrow \Lambda_c\pi^{+}$  & $1$ & $\sqrt{3}/2$
&$\Lambda_{c1}(2593) \rightarrow \Sigma^{++}_c\pi^{-}$ & $1$ & $3/2$ \\
\hline
$\Sigma^{*0}_{c} \rightarrow \Lambda_c\pi^{-}$ & $1$ & $1/2$
&$\Sigma_{c1}^0(2670) \rightarrow \Sigma_c^+\pi^{-}$ & $1$ & $3/2$ \\
\hline
$\Sigma^{* ++}_{c} \rightarrow \Lambda_c\pi^{+}$& $1$ & $1/2$
&$\Xi^{*}_{c1}(2815) \rightarrow \Xi^{*0}_c\pi^{+}$ & $1/\sqrt{2}$ & $1/2$ \\
\hline
$\Xi^{*0}_{c} \rightarrow \Xi^{0}_c\pi^{0}$ & $1/2$ & $1/2$
&$\Xi^{*}_{c1}(2815) \rightarrow \Xi^{*+}_c\pi^{0}$ & $1/2$ & $1/2$ \\
\hline
$\Xi^{*0}_{c}\rightarrow\Xi^{+}_c\pi^{-}$& $1/\sqrt{2}$&$1/2$
&$\Lambda^{*}_{c1}(2625)\rightarrow\Sigma^{0}_c\pi^{+}$ & $1$ & $\sqrt{3}/2$  \\
\hline
$\Xi^{*+}_{c} \rightarrow \Xi^{0}_c\pi^{+} $    & $1/\sqrt{2}$ &  $1/2$
&$\Lambda^{*}_{c1}(2625)\rightarrow\Sigma^{+}_c\pi^{0}$& $1$ &$\sqrt{3}/2$ \\
\hline
$\Xi^{*+}_{c} \rightarrow \Xi^{+}_c\pi^{0}$&1/2&$1/2$
&$\Lambda^{*}_{c1}(2625)\rightarrow\Sigma^{++}_c\pi^{-}$&$1$&$\sqrt{3}/2$\\
\hline
& & & $\Xi^{*}_{c1}(2815)\rightarrow\Xi^{0\prime}_c\pi^{+}$
& $1/\sqrt{2}$ & $\sqrt{3}/2$ \\
\hline
& & & $\Xi^{*}_{c1}(2815) \rightarrow \Xi^{+\prime}_c\pi^{0}$
& $1/2$ & $\sqrt{3}/2$\\
\hline
& & &$\Sigma^{*0}_{c1}(2701) \rightarrow \Sigma^{+}_c\pi^{-}$
& $1$ & $\sqrt{3}/2$\\
\end{tabular}
\end{table}

\begin{table}[t]
\caption{Charm baryon-pion couplings.}
\begin{tabular}{|c|c|c|}
Coupling & Our & Ref. \cite{TOK}\\
\hline\hline
$g_{\Sigma_c\Lambda_c\pi}$ & 8.88 GeV$^{-1}$ & 6.51$\pm$0.35 GeV$^{-1}$   \\
$g_{\Xi_c^*\Xi_c\pi}$ & 8.34 GeV$^{-1}$ & 6.585$\pm$0.375 GeV$^{-1}$        \\
\hline
$f_{\Lambda_{c1; S}\Sigma_c\pi}$ & 0.52 & 0.665$\pm$0.135\\
$f_{\Xi_{c1; S}^*\Xi_c^*\pi}$ & 0.36 & 0.45$\pm$0.13\\
$f_{\Sigma_{c1; A}\Sigma_c\pi}$ & 0.07 & 0.075$\pm$0.015\\
\hline
$f_{\Lambda_{c1; S}^*\Sigma_c\pi}$ & 21.5 GeV$^{-2}$
&50.85$\pm$14.25 GeV$^{-2}$\\
$f_{\Xi_{c1; S}^*\Xi_c^\prime\pi}$ & 20 GeV$^{-2}$ &32.25$\pm$8.15 GeV$^{-2}$\\
$f_{\Sigma_{c1; A}^*\Sigma_c^\prime\pi}$&0.50 GeV$^{-2}$&0.75$\pm$0.15 GeV$^{-2}$\\
\end{tabular}
\end{table}

\begin{table}[t]
\caption{Strong one-pion decay rates.}
 \begin{tabular}{|c|c|c|c|}
 $B_Q\rightarrow B^{\prime}_{Q}\pi$ & Our & Ref. \cite{TOK}  &  Experiment  \\
\hline\hline
\multicolumn{4}{|l|}{P-wave transitions} \\
\hline
$\Sigma^{+}_{c} \rightarrow \Lambda_c\pi^{0} $ & $ 3.63\pm 0.27 $ MeV
& $1.555\pm 0.165$ MeV& $  $\\
$\Sigma^{0}_{c} \rightarrow \Lambda_c\pi^{-} $ & $ 2.65\pm 0.19 $ MeV
& $1.435\pm 0.155$ MeV& \\
$\Sigma^{++}_{c} \rightarrow \Lambda_c\pi^{+}$ & $ 2.85\pm 0.19 $ MeV
& $1.505\pm 0.165$ MeV& $  $\\
\hline
$\Sigma^{*0}_{c} \rightarrow \Lambda_c\pi^{-}$ & $ 21.21 \pm 0.81$ MeV
& $ 11.365 \pm 1.215$ MeV&
$ 13.0^{+3.7}_{-3.0} $ MeV  \\
$\Sigma^{* ++}_{c} \rightarrow \Lambda_c\pi^{+}$ &  $ 21.99\pm 0.87 $ MeV
& $ 11.765\pm 1.265$ MeV&
$ 17.9^{+3.8}_{-3.2} $ MeV  \\
\hline
$\Xi^{*0}_{c} \rightarrow \Xi^{0}_c\pi^{0} $ &  $ 1.01 \pm 0.15$ MeV
& $ 0.525\pm 0.055$ MeV& \\
$\Xi^{*0}_{c} \rightarrow \Xi^{+}_c\pi^{-} $ &  $ 2.11 \pm 0.29$ MeV
& $ 1.30\pm 0.15$ MeV&
$ \Gamma(\Xi^{*0})< 5.5 $ MeV  \\
\hline
$\Xi^{*+}_{c} \rightarrow \Xi^{0}_c\pi^{+} $ &  $ 1.78 \pm 0.33  $ MeV
& $ 1.09 \pm 0.12$ MeV& \\
$\Xi^{*+}_{c} \rightarrow \Xi^{+}_c\pi^{0} $ &  $ 1.26 \pm 0.17  $ MeV
& $ 0.67\pm 0.08$ MeV&
$ \Gamma(\Xi^{*+})< 3.1 $ MeV\\
\hline \hline
\multicolumn{4}{|l|}{S-wave transitions} \\
\hline
$\Lambda_{c1; S}(2593) \rightarrow \Sigma^{0}_c\pi^{+} $ & $0.83\pm 0.09  $ MeV
& $ 1.775\pm 0.695$ MeV
& $0.86^{+0.73}_{-0.56}$ MeV\\
$\Lambda_{c1; S}(2593) \rightarrow \Sigma^{+}_c\pi^{0} $ & $0.98\pm 0.12  $ MeV
& $1.18\pm 0.46$ MeV
& $ \Gamma({\Lambda_{c1; S}}) = 3.6^{+2.0}_{-1.3}$ MeV \\
$\Lambda_{c1; S}(2593) \rightarrow \Sigma^{++}_c\pi^{-} $ & $0.79\pm 0.09$ MeV&
$ 1.47\pm 0.57$ MeV
& $0.86^{+0.73}_{-0.56}$ MeV\\
\hline
$\Xi^{*}_{c1; S}(2815) \rightarrow \Xi^{*0}_c\pi^{+} $ &  $ 0.46\pm 0.03$  MeV
& $ 1.575\pm 0.835$ MeV
& \\
$\Xi^{*}_{c1; S}(2815) \rightarrow \Xi^{*+}_c\pi^{0} $ &  $ 0.24\pm 0.02$ MeV
& $0.775\pm 0.415$ MeV
& $ \Gamma({\Xi^{*}_{c1; S}})<2.4 $ MeV \\
\hline
$\Sigma^{0}_{c1; A}(2670) \rightarrow \Sigma^{+}_c\pi^{-} $
& $0.113\pm 0.001$ MeV
& $ 0.137\pm 0.052$ MeV& \\
\hline \hline
\multicolumn{4}{|l|}{D-wave transitions} \\
\hline
$\Lambda^{*}_{c1; S}(2625) \rightarrow \Sigma^{0}_c\pi^{+} $
& $ 0.080\pm 0.009 $ MeV
& $0.465\pm 0.245$ MeV& $< 0.13$ MeV\\
$\Lambda^{*}_{c1; S}(2625) \rightarrow \Sigma^{+}_c\pi^{0} $
& $ 0.095\pm 0.012 $ MeV
& $ 0.42\pm 0.22$ MeV& $\Gamma({\Lambda^{*}_{c1}}) <1.9 $MeV  \\
$\Lambda^{*}_{c1; S}(2625) \rightarrow \Sigma^{++}_c\pi^{-} $
& $ 0.076\pm 0.009 $ MeV
& $ 0.44\pm 0.23$ MeV& $<0.15$ MeV\\
\hline
$\Xi^{*}_{c1; S}(2815) \rightarrow \Xi^{0\prime}_c\pi^{+}$
& $0.35\pm 0.05$ MeV
& $ 0.84\pm 0.43$ MeV& $ $  \\
$\Xi^{*}_{c1; S}(2815) \rightarrow \Xi^{+\prime}_c\pi^{0} $
& $0.21\pm 0.03$ MeV
& $0.435\pm 0.205$ MeV& $\Gamma({\Xi_{c1}^*}) < 2.4$ MeV\\
\hline
$\Sigma^{*0}_{c1; A}(2701) \rightarrow \Sigma^{+}_c\pi^{-} $
& $0.0012\pm 0.0001$ MeV
& $ 0.0025\pm 0.0015$ MeV& \\
\end{tabular}
\end{table}

\begin{table}[t]
\caption{Radiative decay rates}
\begin{tabular}{|c|c|c|c|}
$B_Q\rightarrow B^{\prime}_{Q}\gamma$ & This approach & Other approaches &
Experiment \cite{PDG}\\
\hline
$\Sigma^{+}_{c}\rightarrow \Lambda_c^+\gamma$ & $ 60.7\pm 1.5$ KeV
& $93$ KeV \cite{Cheng1} & $  $\\
\hline
$\Sigma^{*+}_{c}\rightarrow\Lambda_c^+\gamma$ & $ 151 \pm 4$ KeV& & \\
\hline
$\Sigma^{*+}_{c}\rightarrow\Sigma_c^+\gamma$ & $ 0.14 \pm 0.004$ KeV& & \\
\hline
$\Xi^{'+}_{c}\rightarrow \Xi_c^+\gamma$  &$ 12.7\pm 1.5$ KeV
& $16$ KeV \cite{Cheng1} &\\
\hline
$\Xi^{'0}_{c}\rightarrow\Xi_c^0\gamma$&$0.17\pm 0.02$ KeV
&$0.3$ KeV \cite{Cheng1}&\\
\hline
$\Xi^{*+}_{c}\rightarrow\Xi_c^+\gamma$&$54\pm 3$ KeV & &\\
\hline
$\Xi^{*0}_{c}\rightarrow\Xi_c^0\gamma$&$0.68 \pm 0.04$ KeV  & & \\
\hline
$\Lambda_{c1; S}(2593) \rightarrow \Lambda_c^+\gamma $ & $0.115\pm 0.001$ MeV
& $0.191c^2_{RT}$ MeV \cite{Cho} & $< 2.36^{+1.31}_{-0.85}$ MeV \\
 & & $0.016$ MeV \cite{Chow} &\\
\hline
$\Lambda_{c1; S}(2593) \rightarrow \Sigma_c^+\gamma $ & $0.077\pm 0.001$ MeV
& $0.127c^2_{RS}$ \cite{Cho} & \\
\hline
$\Lambda_{c1; S}(2593) \rightarrow \Sigma_c^{*+}\gamma $ & $0.006\pm 0.0001$ 
MeV & $0.006c^2_{RS}$ \cite{Cho} & \\
\hline
$\Lambda_{c1; S}^*(2625) \rightarrow \Lambda_c^+\gamma$&$0.151\pm 0.002$ MeV
& $0.253c^2_{RT}$ MeV \cite{Cho} & $< 1$ MeV\\
 & & $0.021$ MeV \cite{Chow} &\\
\hline 
$\Lambda_{c1; S}^*(2625) \rightarrow \Sigma_c^+\gamma $ & $0.035\pm 0.0005$ MeV
& $0.058c^2_{RS}$ \cite{Cho} & \\
\hline
$\Lambda_{c1; S}^*(2625) \rightarrow \Sigma_c^{*+}\gamma $ 
& $0.046\pm 0.0006$ MeV & $0.054c^2_{RS}$ \cite{Cho} & \\
\hline
$\Xi_{c1; S}^{*+}(2815) \rightarrow \Xi_c^+\gamma $ & $0.190\pm 0.005$ MeV& &\\
\hline
$\Xi_{c1; S}^{*0}(2815) \rightarrow \Xi_c^0\gamma $ & $0.497\pm 0.014$ MeV& &\\
\hline
$\Lambda_{b1; S}(5933)
\rightarrow \Lambda_b^0\gamma $ & $0.128\pm 0.022$ MeV
& $0.09$ MeV \cite{Chow}& \\
\hline
$\Lambda_{b1; S}^*(5966)
\rightarrow \Lambda_b^0\gamma $ & $0.172\pm 0.026$ MeV
& $0.119$ MeV \cite{Chow}& \\
\end{tabular}
\end{table}

\newpage


\begin{figure}[t]
\centering{\
\epsfig{figure=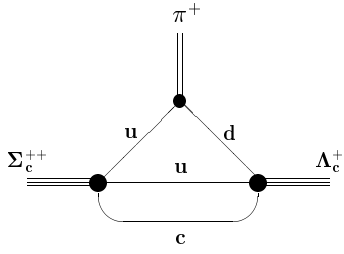,height=3.5cm}}
\caption{Triangle diagram with a pion emitted by light quark.}
\end{figure}
\vspace{1cm}

\begin{figure}[t]
\centering{\
\epsfig{figure=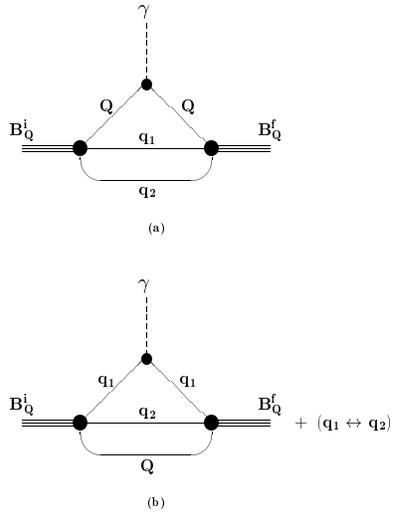,height=7cm}}
\caption{Triangle diagrams with a photon emitted by (a) heavy and
(b) light quark.}
\end{figure}
\newpage
\begin{figure}[t]
\centering{\
\epsfig{figure=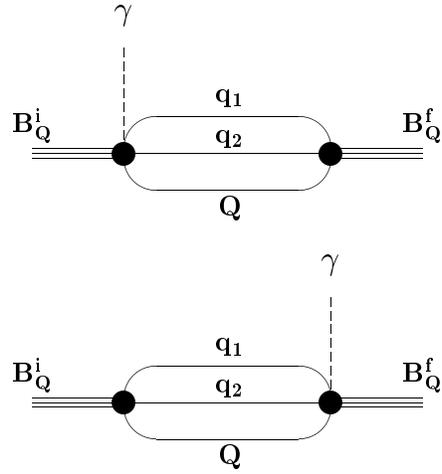,height=7cm}}
\caption{Contact interaction-type diagrams.}
\end{figure}
\vspace{1cm}

\begin{figure}[t]
\centering{\
\epsfig{figure=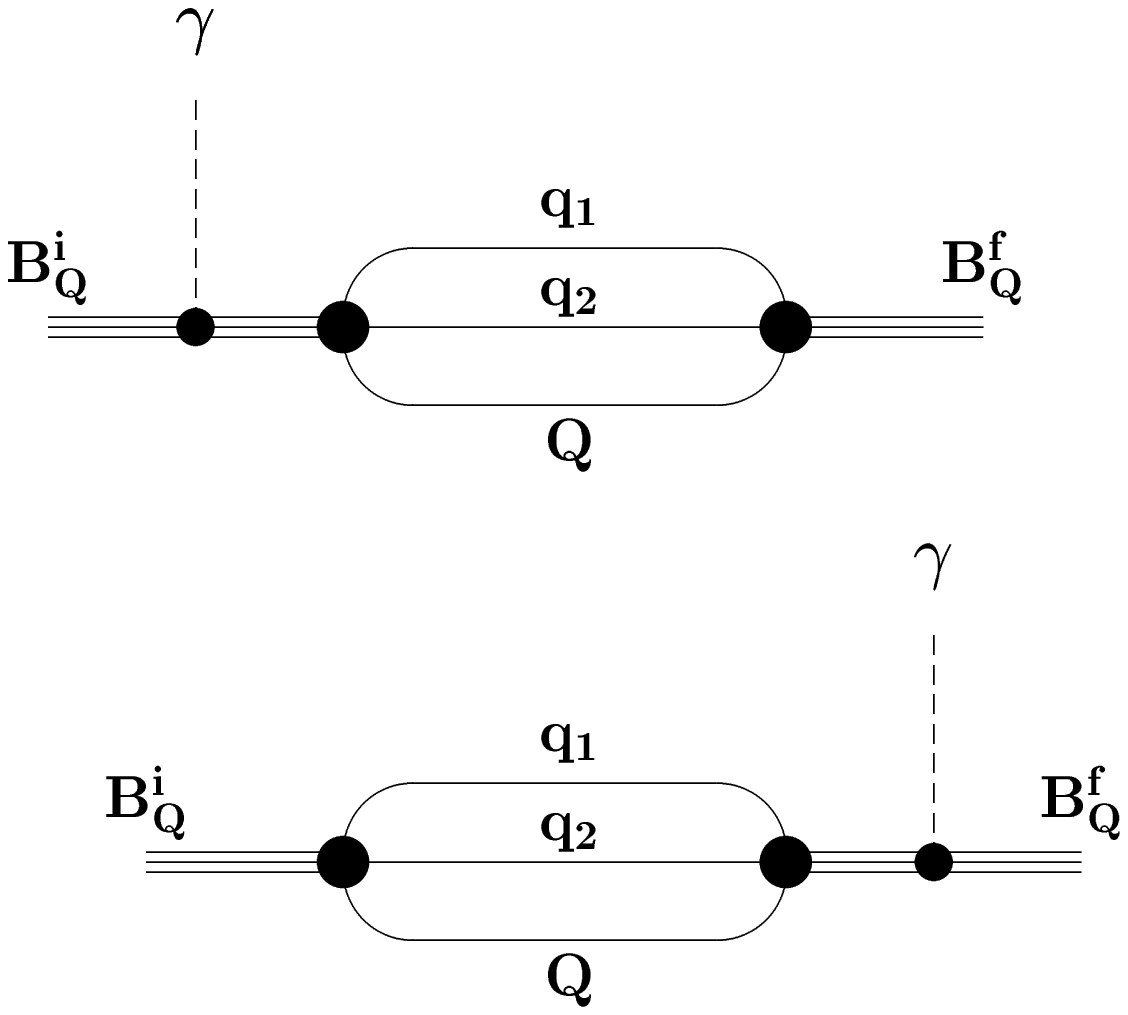,height=7cm}}
\caption{Pole diagrams.}
\end{figure}

\begin{thebibliography}{999}
%
%
\bibitem{PDG} C. Caso et.al. (Particle Data Group),
Eur. Phys. J.  {\bf C3} (1998) 1.
%
\bibitem{CLEO1} G. Brandenburg et.al., CLEO Coll.,
Phys. Rev. Lett {\bf 78}, 2304 (1997).
%
\bibitem{CLEO2} P. Avery el.al., CLEO Coll., Phys. Rev. Lett.
{\bf 75}, 4364 (1995).
\bibitem{CLEO3} L. Gibbons et.al., CLEO Coll.,
Phys. Rev. Lett. {\bf 77}, 810 (1996).
%
\bibitem{CLEO4} T.E. Coan et al., CLEO Coll., Preprint CLEO CONF 97-29, 1997;\\
P. Jessop et al.,
Phys. Rev. Lett. {\bf 82}, 492 (1999).
%
\bibitem{ARGUS} H. Albrecht et.al., ARGUS Coll., Phys. Lett. {\bf B317},
227 (1993); \\
Phys. Lett. {\bf B402}, 207 (1997).
\bibitem{E687} P.L. Frabetti et.al, E687 Coll.,
Phys. Rev. Lett. {\bf 72}, 961 (1994);\\
Phys. Lett B {\bf 365}, 461 (1996).
\bibitem{CLEO5} K.W. Edwards et.al., CLEO Coll.,
Phys. Rev. Lett. {\bf 74}, 3331 (1995).
%
\bibitem{CLEO6} G. Brandenburg et al., CLEO Coll.,
Preprint CLEO CONF 97-17, 1997.
%
%
\bibitem{KKP} J.G. K\"orner, M. Kr\"amer, and D. Pirjol,
Prog. in Part. Nucl. Phys. {\bf 33}, 787 (1994).
\bibitem{HKLT} F. Hussain, J.G. K\"orner, J. Landgraf, and S. Tawfiq
Z. Phys. {\bf C69}, 655 (1996).
%
\bibitem{HKT} F. Hussain, J.G. K\"orner, and S. Tawfiq,
Preprints MZ-TH/96-10, IC/96/35, 1996.
\bibitem{PY} D. Pirjol and T.M. Yan, Phys. Rev. D {\bf 56}, 5483 (1997).
\bibitem{TOK} S. Tawfiq, P.J. O'Donnell, and J.G. K\"{o}rner,
Phys. Rev. {\bf D58}, 054010 (1998); \\
Preprint UTPT-98-08, 1998.
%
\bibitem{Yan} T.M. Yan et. al., Phys. Rev. {\bf D46}, 1148 (1992).
\bibitem{Cheng1} H.-Y. Cheng et al. Phys. Rev. {\bf D47}, 1030 (1993).
\bibitem{Cho} P. Cho, Phys. Rev. {\bf D50}, 3295 (1994).
\bibitem{Cheng2} H.-Y. Cheng, Phys. Lett. {\bf B399}, 281 (1997).
\bibitem{HDH} M.-Q. Huang, Y.-B. Dai, and C.-S. Huang,
Phys. Rev. {\bf D52}, 3986 (1995).
%
\bibitem{Chow} C.-K. Chow, Phys. Rev. {\bf D54}, 3374 (1996).
%
\bibitem{CF}G. Chiladze and A. Falk, Phys. Rev. {\bf D56}, 6738 (1997).
%
\bibitem{Savage} M.J.Savage, Phys. Lett. {\bf B345}, 61 (1995).
%
\bibitem{GY} A.G. Grozin and O.I. Yakovlev, Eur. Phys. J. {\bf C2}, 721 (1998).
\bibitem{ZD} S.-L. Zhu and Y.-B. Dai, hep-ph/9810243.
%
\bibitem{EI} G.V. Efimov and M.A. Ivanov, "The Quark Confinement Model",
IOP, 1993.
\bibitem{AIKL} I.V. Anikin, M.A. Ivanov, N.B. Kulimanova and V.E. Lyubovitskij,\\
Z. Phys. C {\bf 65}, 681 (1995); Phys. Atom. Nucl. {\bf 57}, 1021 (1994).
\bibitem{IL} M.A. Ivanov and V.E. Lyubovitskij,
Phys. Lett. {\bf B408}, 435 (1997).
\bibitem{ILL}M.A. Ivanov, M.P. Locher, V.E. Lyubovitskij,
Few-Body Syst. {\bf 21}, 131 (1996).
%
\bibitem{IS} M.A. Ivanov and P. Santorelli, hep-ph/9903446.
\bibitem{SW} A. Salam, Nuovo Cim. {\bf 25}, 224 (1962);
S. Weinberg, Phys. Rev. {\bf 130}, 776 (1963).
%
\bibitem{ILKK} M.A. Ivanov, V.E. Lyubovitskij, J.G. K\"{o}rner and P. Kroll,
Phys. Rev. {\bf D56}, 348 (1997).
\bibitem{IKLR1} M.A. Ivanov, J.G. K\"{o}rner, V.E. Lyubovitskij,
and A.G. Rusetsky,\\
Phys. Rev. {\bf D57}, 5632 (1998); Mod. Phys. Lett. {\bf A13}, 181 (1998).
%
\bibitem{IKLR2} M.A. Ivanov, J.G. K\"{o}rner, V.E. Lyubovitskij,
A.G. Rusetsky, \\
Phys. Lett. {\bf B442}, 435 (1998).
%
\bibitem{IKL} M.A. Ivanov, J.G. K\"{o}rner, V.E. Lyubovitskij,
Phys. Lett. {\bf B448}, 143 (1999).
%
\bibitem{Mandelstam} S. Mandelstam, Ann. Phys. {\bf 19}, 1 (1962).
\bibitem{Terning} J. Terning, Phys. Rev. {\bf D44}, 887 (1991).
%
\bibitem{IKMR} M.A. Ivanov, Yu.L. Kalinovsky, P. Maris, and C.D. Roberts, \\
Phys. Rev. {\bf C57}, 1991 (1998); Phys. Lett. {\bf B416}, 29 (1998).
%
\end{thebibliography}
\end{document}